\documentclass[a4paper,12pt]{article}
\usepackage{jheppub}
\usepackage{epsf,epsfig,subfigure,mathtools,hhline,float,multirow,nicefrac,slashed,color,url,graphicx,subfigure}
\usepackage[utf8x]{inputenc}
\usepackage[section]{placeins}
\usepackage[outdir=./]{epstopdf}
\newcommand{\be}{\begin{equation}}
\newcommand{\ee}{\end{equation}}
\newcommand{\bes}{\begin{equation*}}
\newcommand{\ees}{\end{equation*}}
\newcommand{\bea}{\begin{eqnarray}}
\newcommand{\eea}{\end{eqnarray}}
\newcommand{\beas}{\begin{eqnarray*}}
\newcommand{\eeas}{\end{eqnarray*}}

\newcommand{\tb}{\tan\beta~}

\newcommand{\sba}{\sin(\beta-\alpha)}
\newcommand{\cba}{\cos(\beta-\alpha)}
\newcommand{\ztwo}{\mathbb{Z}_2}

\title{Exploring Wrong Sign Scenarios in the Yukawa-Aligned 2HDM}
\author{Shinya Kanemura,}
\author{Tanmoy Mondal and}
\author{Kei Yagyu}
\affiliation{Department of Physics, Osaka University, Toyonaka, Osaka 560-0043, Japan}
\emailAdd{kanemu@het.phys.sci.osaka-u.ac.jp}
\emailAdd{tanmoy@het.phys.sci.osaka-u.ac.jp}
\emailAdd{yagyu@het.phys.sci.osaka-u.ac.jp}

\abstract{
We discuss scenarios with wrong-sign (WS) Yukawa couplings for the discovered Higgs boson
in the Yukawa-aligned two Higgs doublet model. In the WS scenario, Yukawa couplings for
down-type quarks and/or charged leptons have an opposite sign as compared to those of the
Higgs boson in the standard model, which can be consistent with current flavor data and
the Higgs signal strengths. 
The phenomenology of additional Higgs bosons in such a scenario can be significantly
 different from that with right-sign Yukawa couplings, mainly due
to a larger Higgs boson mixing to be required in the wrong-sign case.
We show the parameter space which is excluded or explored by direct searches for the
additional Higgs bosons at the current and high-luminosity LHC under the constraints from
 perturbative unitarity and vacuum stability. In particular, we find that most of the
parameter space is explored in the WS scenario with the Type-X (lepton specific) Yukawa
 interaction which is a special case of the Yukawa
alignment realized by imposing a softly-broken $\ztwo$ symmetry.
We propose that multi-Higgs events from pair productions of the additional Higgs
bosons can be the smoking gun signature to probe the WS scenario, and give the expected number of
events at the high-luminosity LHC.
}

\preprint{OU-HET 1156}
\date{}
\begin{document}

\maketitle

\section{Introduction}
Current LHC data show that the properties of the discovered Higgs boson are consistent with
those in the Standard Model (SM)~\cite{ATLAS:2020qdt,CMS:2020gsy}.
On the other hand, the Higgs sector is extended from the minimal form with one isospin
doublet field in various new physics models, which are motivated to explain phenomena
beyond the SM, such as neutrino oscillations, dark matter and baryon asymmetry of the
Universe.
In an extended Higgs sector, the so-called Higgs alignment is approximately required
from the above situation, where one of the mass eigenstates of neutral Higgs bosons with
a mass of 125 GeV ($h$) coincides with a component of the Higgs doublet field whose vacuum
expectation value (VEV) takes the SM value. In the Higgs alignment limit, all the Higgs
boson couplings with SM particles take the same values as those in the SM at the tree
level. Thus, the Higgs alignment is now getting quite important in models with extended
Higgs sectors to accommodate the current situation.

The Higgs alignment is realized by taking the decoupling limit, in which all the masses
of additional Higgs bosons are taken to be infinity. This, however, is essentially the
same as the SM-limit, so that desired properties of extended Higgs sectors, e.g.,
explaining unsolved phenomena and testability at collider experiments, must be spoiled.
There is an alternative and attractive way to realize the Higgs alignment, i.e., the
so-called alignment without decoupling, in which mixing of $h$ with the other neutral
states vanishes. In this case, masses of additional Higgs bosons can be at the electroweak
scale, while couplings of $h$ are kept at the SM values. In practice, we do not need to
demand the exact Higgs alignment because the observed $h$ couplings, defined by the
$\kappa$-scheme~\cite{LHCHiggsCrossSectionWorkingGroup:2013rie}, include typically of
order 10\% uncertainties. In the nearly aligned scenario, the Higgs boson coupling to the
weak bosons takes a similar value to the SM one. In contrast, particularly in models with
multi-Higgs doublets, Yukawa couplings can have either right-sign (RS) or wrong-sign (WS)
with a similar magnitude to those in the SM. In the RS (WS) scenario, the sign of Yukawa
couplings is the same (opposite) as compared to the corresponding SM ones. Thus,
measurement of the ``sign''\footnote{To be more precise, we here mean the relative sign 
between a fermion mass and a coefficient of a Higgs-fermion-antifermion vertex. } of Yukawa 
couplings is crucial to understand the structure of the Higgs sector.

The WS scenario, having WS Yukawa couplings, has been studied earlier in
Refs.~\cite{Ginzburg:2001ss,Ferreira:2014naa,Dumont:2014wha,Fontes:2014tga,Biswas:2015zgk,Modak:2016cdm,Ferreira:2017bnx,Su:2019ibd,Han:2020zqg,Raju:2020hpe} in the Type-II two Higgs doublet
model (2HDM), where Yukawa couplings for down-type quarks and charged leptons are WS.
In this case, most of the parameter space has already been excluded by the constraints
from $B$ physics data, the Higgs signal strength and direct searches for additional Higgs
bosons~\cite{Su:2019ibd}.
The WS scenario has also been discussed in the
Type-X 2HDM based on the motivation to explain the anomaly of the muon anomalous magnetic moment
$(g-2)_\mu$~\cite{Abe:2015oca,Cherchiglia:2017uwv,Wang:2018hnw}, where a smaller mass of the CP-odd Higgs boson $(A)$ than half of
the mass of $h$ is required. 
See also Refs.~\cite{Cao:2009as,Broggio:2014mna} (Ref.~\cite{Ilisie:2015tra}) for the discussion of the $(g-2)_\mu$ anomaly in the Type-X 2HDM (Yukawa aligned 2HDM defined below) with the light CP-odd Higgs boson.
Although such a light $A$ can easily be excluded by the
constraint $\text{BR}(h\to AA) \lesssim 5\%$ with leptonic decays of $A$ given
at the LHC~\cite{ATLAS:2015unc,CMS:2017dmg,ATLAS:2018coo,CMS:2018qvj,CMS:2019spf,CMS:2020ffa},
the WS scenario in the Type-X 2HDM allows to take the $hAA$ coupling to be
negligibly small~\cite{Abe:2015oca,Chun:2016hzs,Cherchiglia:2017uwv,Chun:2017yob,Ghosh:2021jeg,Kim:2022xuo,Wang:2018hnw}. Unlike
the Type-II 2HDM, the phenomenology of the additional Higgs bosons has not intensively
been discussed, since gluon fusion production is inefficient due to the significant
suppression of quark Yukawa couplings in the Type-X 2HDM.

So far, the WS scenario has been studied in the 2HDMs with a softly-broken $\ztwo$
symmetry, for e.g., the Type-II and Type-X 2HDMs as aforementioned. In this work, we focus
on the WS scenario in Yukawa aligned 2HDMs (A2HDMs), where flavor
changing neutral currents (FCNCs) via Higgs boson exchanges at the tree level vanish
due to the assumption that one of the Yukawa matrices for each charged fermion is
proportional to the other~\cite{Pich:2009sp}. 

The A2HDM provides additional CP-violating (CPV) phases in
Yukawa interactions and the Higgs potential, which do not appear in the 2HDMs with the
$\ztwo$ symmetry. Recently, it has been found that severe constraints from electric dipole
moments can be avoided due to destructive interferences among fermion and scalar
loop diagrams~\cite{Kanemura:2020ibp}, and the testability of the CPV phases at future
lepton colliders have been discussed in Ref.~\cite{Kanemura:2021atq}. Furthermore, in
Refs.~\cite{Enomoto:2021dkl,Enomoto:2022rrl}, it has been clarified that the observed
baryon asymmetry of the Universe can be explained based on the electroweak baryogenesis
scenario in the CPV A2HDM.

In this paper, we investigate the CP-conserving A2HDM\footnote{See
Ref.~\cite{Eberhardt:2020dat} for a recent global fit in the A2HDM.} with WS Yukawa
couplings under the constraints coming from perturbative unitarity~\cite{Kanemura:1993hm,Akeroyd:2000wc,Horejsi:2005da,Ginzburg:2005dt,Kanemura:2015ska}, vacuum stability~\cite{Deshpande:1977rw,Nie:1998yn,Kanemura:1999xf},
flavor data, the Higgs signal strength and the direct searches at the LHC.
As a special case of the A2HDM, we also explore the Type-X 2HDM without requiring the
light $A$, which has been studied in the previous works mentioned above. We show that in
the WS scenario, bosonic decay channels of the additional Higgs bosons, such as $H \to
hh/WW/ZZ$, $A \to Zh$, $H^ \pm \to W^ \pm h$~\footnote{See
Refs.~\cite{Kanemura:2022ldq}, \cite{Aiko:2022gmz} and \cite{Aiko:2021can} for the recent
study on electroweak radiative corrections to the $H$, $A$ and $H^ \pm $ decays in the
2HDMs with the $\ztwo$ symmetry, respectively.} tend to be more important as compared with
those in the RS scenario.
We find that the direct searches for additional Higgs
bosons at the High-Luminosity LHC (HL-LHC) can explore most of the parameter
space in the Type-X scenario, which is allowed by the theoretical constraints. 
In addition, in the A2HDM, the HL-LHC can explore a large portion of the parameter space, up to 
around 1 TeV for the masses of additional Higgs
bosons. We propose that multi-Higgs signatures from pair productions of the additional
Higgs boson can be the smoking gun to probe the WS scenario and give the expected number
of events at the HL-LHC.

This paper is organized as follows: We begin with a brief discussion of the A2HDM and the
structure of Yukawa interactions in Sec.~\ref{sec:model}. Next, in
Sec.~\ref{sec:constraints}, we discuss various theoretical and experimental constraints
and their impact on the WS scenario. To distinguish the WS and RS scenarios, in
Sec.~\ref{sec:comparison}, we compare the branching ratios of additional neutral Higgs
bosons. Sec.~\ref{sec:LHC-limit} shows the current limit on the parameter space in the
A2HDM from direct searches at the LHC Run-II experiment, In Sec.~\ref{sec:proposal}, we discuss the
prospect of looking for the WS scenarios via multi-Higgs searches at the HL-LHC.
We then conclude our findings in Sec.~\ref{sec:conclusion}.


\section{WS Yukawa Couplings in 2HDMs}\label{sec:model}

The 2HDM consists of two isospin scalar doublets $\Phi_1$ and $\Phi_2$ with
hypercharge $Y = 1/2$. For reviews of the 2HDM, see~e.g., \cite{Gunion:1989we,Djouadi:2005gj,Branco:2011iw}. 
Throughout this paper, we neglect new CPV phases in the Higgs sector for simplicity. 

We first define the Higgs basis~\cite{Botella:1994cs,Davidson:2005cw} expressed as 
\begin{align}
\begin{pmatrix}
\Phi_1 \\
\Phi_2
\end{pmatrix}
= 
\begin{pmatrix}
1 & 0 \\
0 & e^{i\xi}
\end{pmatrix}
\begin{pmatrix}
\cos\beta & -\sin\beta \\
\sin\beta & \cos\beta
\end{pmatrix}
\begin{pmatrix}
\Phi \\
\Phi'
\end{pmatrix}, 
\end{align}
where $\xi$ is the relative phase of two VEVs. 
In the Higgs basis, the doublets are parametrized as,
\be
\Phi=\begin{pmatrix}  G^ \pm \\\frac{1}{\sqrt2}(v+S_1+iG^0)\end{pmatrix},\quad
\Phi'=\begin{pmatrix}  H^ \pm \\\frac{1}{\sqrt2}(S_2+i A)\end{pmatrix}, 
\ee
where only $\Phi$ contains the VEV $v \simeq 246$ GeV.  
In the above expression, $G^\pm$ $(G^0)$ are the Nambu-Goldstone bosons which are absorbed 
into the longitudinal component of the $W^\pm$ $(Z)$ bosons, while 
$H^\pm$, $A$ and $S_{1,2}$ are respectively the physical singly-charged, CP-odd and CP-even Higgs states. 

The general scalar potential is given by,
\begin{align}
\nonumber V &= m^2 |\Phi|^2 + M^2|\Phi'|^2 - (\mu^2\Phi^{\dagger}\Phi' + \text{h.c.})
+\frac{\Lambda_1}{2}|\Phi|^4 +\frac{\Lambda_2}{2}|\Phi'|^4+\Lambda_3|\Phi|^2|\Phi'|^2 +\Lambda_4|\Phi^\dagger\Phi'|^2 \notag\\
& +\left[\frac{\Lambda_5}{2}(\Phi^\dagger\Phi') + \Lambda_6|\Phi|^2 + \Lambda_7|\Phi'|^2\right](\Phi^\dagger\Phi') + \text{h.c.},
\label{eq:2hdm-pot}
\end{align}
where $\mu^2$ and $\Lambda_{5,6,7}$ are assumed to be real. 
Minimization of the potential gives the following tadpole solutions,
\be
m^2 = -\frac{\Lambda_1}{2} v^2 \quad, \quad \mu^2 = \frac{\Lambda_6}{2} v^2.
\ee
To obtain the mass eigenstates of the CP-even states, 
we introduce the mixing angle $\beta-\alpha$ as 
\footnote{In the general 2HDM, the parameter $\beta$ does not have physical impacts, 
but here we express the mixing angle to be $\beta - \alpha$
as the analogy to the 2HDMs with the softly-broken $Z_2$ symmetry. },
\begin{equation}
\begin{pmatrix}  H\\h
\end{pmatrix}=
\begin{pmatrix}  \cba&-\sba\\ \sba&\cba
\end{pmatrix}
\begin{pmatrix}  S_1\\S_2
\end{pmatrix}, 
\end{equation}
where we can identify $h$ with the discovered Higgs boson with the mass of 125 GeV. 
Some of the parameters in the potential can be rewritten in terms of masses and mixing angles as follows:
\bea
\Lambda_1&=&\frac{\sin ^2(\beta-\alpha ) M_h^2+\cos ^2(\beta-\alpha ) M_H^2}{v^2}, \\
\Lambda_3&=& \frac{2(M_{H^ \pm }^2 - M^2)}{v^2}, \\
\Lambda_4&=&\frac{\cos^2(\beta- \alpha ) M_h^2 + \sin^2(\beta- \alpha)M_H^2+M_A^2 -2M_{H^ \pm }^2 }{v^2}, \\
\Lambda_5&=&\frac{\cos ^2(\beta- \alpha ) M_h^2+\sin ^2(\beta- \alpha )M_H^2-M_A^2}{v^2}, \\
\Lambda_6&=&\frac{\sin (\beta- \alpha ) \cos (\beta -\alpha )
\left(M_h^2-M_H^2\right)}{v^2}. \label{eq:lam6}
\eea
For our phenomenological analysis, we choose the following set of input parameters:
\begin{align}
v,~M_h,~M_H,~M_A,~M_{H^ \pm },~\Lambda_2,~\Lambda_3,~\Lambda_7,~\text{ and } \cba, 
\end{align}
where $v$ and $M_h$ are fixed to be about 246 GeV and 125 GeV, respectively. 

We note in passing that in the 2HDMs with the $\ztwo$ symmetry, 
two of seven quartic couplings, i.e.,  the coefficients of $|\Phi_1|^2(\Phi_1\Phi_2^\dagger)$ 
and $|\Phi_2|^2(\Phi_1\Phi_2^\dagger)$ in the general basis, 
are forbidden. 
As a result, two quartic couplings in the Higgs basis are written by the other parameters. 
For instance, $\Lambda_2$ and $\Lambda_7$ are written as 
\begin{align}
\Lambda_2   & = \Lambda_1\left(3 - \frac{2}{\sin^22\beta}\right) + 2(\Lambda_3 +\Lambda_4 + \Lambda_5)\cot^22\beta -\frac{2\Lambda_6}{\sin^3 2\beta}(\cos 2\beta + \cos 6\beta), \label{eq:lam237} \\
\Lambda_7   & = \Lambda_6(1 -2\cot^22\beta) + (\Lambda_3 +\Lambda_4 + \Lambda_5
-\Lambda_1)\cot2\beta.  \label{eq:lam7}
\end{align}
These relations turn out to be important when we consider the constraint on the parameter 
space from perturbative unitarity as it will be discussed in Sec.~\ref{sec:constraints}. 

The Yukawa interactions are generally expressed in the mass eigenstates of fermions as,
\begin{align}
\mathcal{L}_Y &=  - \bar{Q}_L^d \left(\frac{\sqrt{2}M_d}{v} \Phi + \rho_d\Phi'\right)d_R - \bar{Q}_L^u \left(\frac{\sqrt{2}M_u}{v} \tilde{\Phi} + \rho_u\tilde{\Phi}'\right)u_R \notag\\
&- \bar{L}_L \left(\frac{\sqrt{2}M_e}{v}\Phi + \rho_e\Phi' \right)e_R + \text{h.c.}, 
\end{align}
where $\tilde{\Phi} = i\sigma_2\Phi^*$, $Q_L^u = (u_L,V_{\textrm{CKM}}d_L)^T$, $Q_L^d = (V_{\textrm{CKM}}^\dagger u_L,d_L)^T$ and $L_L = (\nu_L,e_L)$ with $
V_{\textrm{CKM}}$ being the Cabibbo-Kobayashi-Maskawa matrix. 
In the above expression, $M_f$ ($f = u,d,e$) are the diagonal mass matrices for charged fermions, 
and $\rho_f$ are arbitrary  $3\times3$ complex matrices which can cause FCNCs mediated by the Higgs bosons at tree level. 
In order to avoid such FCNCs, we impose the Yukawa-alignment~\cite{Pich:2009sp} in the flavor space: \footnote{The Yukawa alignment condition is not protected by a symmetry, so that it is generally broken 
at loop levels. The deviation from the alignment condition has been studied by using the renormalization group equations in Refs.~\cite{Kanemura:2020ibp,Gori:2017qwg}, and
found that the size of the deviation is typically quite small as this should be proportional to small off-diagonal components of the CKM matrix. Throughout this paper, 
we neglect such a loop effect on the Yukawa couplings. }
 \be
\rho_{d,e} = \sqrt{2}\zeta_{d,e}\frac{M_{d,e}}{v} \quad \textrm{and }\quad \rho_u = \sqrt{2}\zeta_u^\ast ~\frac{M_u}{v}.
 \ee
We call $\zeta_f$ the alignment parameters which are generally complex-valued. After
applying the aforementioned alignment condition, we obtain the Yukawa interaction terms for the physical Higgs bosons as 
follows:
 \bea\label{eq:yuk-mass-basis}
 \mathcal{L}_{Y}^{\rm int} &=&  -\sum_{f=u,d,e} \bar f_L\dfrac{M_f}{v} f_R\left( \kappa_f  \,h +
\kappa_f^H  \,H + i \,\kappa_f^A \, A\right) \nonumber\\
 && - \dfrac{\sqrt2}{v}H^+\left(\zeta_{d} \bar u_L V_{\textrm{CKM}} M_{d}  d_R -\zeta_{u} \bar u_R  M_{u}^\dagger V_{\textrm{CKM}} d_L + \zeta_{e} \bar\nu_L  M_{e}  e_R  \right)  + \text{h.c.} 
 \eea
The Yukawa coupling modifiers are given by 
\bea\label{eq:kappa}
\kappa_f &=& \sba + \zeta_f \cba, \nonumber \\  
\kappa_f^H &=& \cba - \zeta_f \sba, \quad
\textrm{and}\quad  \kappa_f^A = -2 I_{3f}\zeta_f, 
\eea
with $I_{3f}$ being the third component of the isospin. 

Apart from the Yukawa alignment, the Higgs-mediated FCNCs can also be forbidden by
imposing the softly-broken $\ztwo$ symmetry, which restricts each type of the right-handed fermions to
couple to only one of $\Phi_1$ and $\Phi_2$. Depending on the $\ztwo$ charge assignment, four different
types of Yukawa interactions are possible~\cite{Barger:2009me,Grossman:1994jb,Aoki:2009ha}. 
The $\ztwo$ symmetric scenarios are the special case of the A2HDMs, i.e., 
the alignment parameters $\zeta_f$ are expressed by the single parameter $\beta$ as 
\begin{align}
\zeta_u &= \zeta_d = \zeta_e = 1/\tan\beta~~(\text{Type-I}) , \notag\\
\zeta_u &= 1/\tb\quad\textrm{and}\quad \zeta_d = \zeta_e = -\tan\beta~~(\text{Type-II}), \notag\\
\zeta_u &= \zeta_d = 1/\tb\textrm{and}\quad\zeta_e = - \tb~~(\text{Type-X}), \notag\\
\zeta_u &= \zeta_e = 1/\tb\textrm{and}\quad\zeta_d = - \tb~~(\text{Type-Y}). \label{eq:types}
\end{align}

The scaling factor $\kappa_f$ in Eq.~(\ref{eq:yuk-mass-basis}) parameterizes the deviation in the Yukawa
couplings from the SM values. 
Similarly, we can define $\kappa_V^{} = \sin(\beta-\alpha)$ for the weak bosons. 
Current measurements restrict $\kappa$ values close to unity. 
From Eq.~(\ref{eq:kappa}), we see that $\kappa_f$ stays close to the SM values when $\sba\simeq 1$ with 
$|\zeta_f\cos(\beta-\alpha)| \ll 1$, which is called the RS scenario. 
It is also possible to have $\kappa_f =  -1$ when 
\begin{align}
\zeta_f = -\frac{1}{\cos(\beta-\alpha)}[1 + \sin(\beta-\alpha)], \label{eq:ws_limit}
\end{align}
which is approximately expressed as $\zeta_f \sim -2/\cba$ for $\sba \simeq 1$. 
We call the limit defined in Eq.~(\ref{eq:ws_limit}) as the WS limit for the Yukawa coupling
\footnote{Notice here that 
by substituting $\zeta_f = -\tan\beta$ in Eq.~(\ref{eq:ws_limit}) we obtain the WS limit $\sin(\beta+\alpha)=1$~\cite{Ferreira:2014naa} for the $\mathbb{Z}_2$ symmetric case.}
In the following, we discuss the parameter space which can give rise to the approximate WS limit, and study relevant constraints on such a parameter space.

\section{Constraints on WS Yukawa Couplings}\label{sec:constraints}

\subsection{Higgs Signal Strengths}

\begin{figure}[t]
\begin{center}
\includegraphics[width=8cm ]{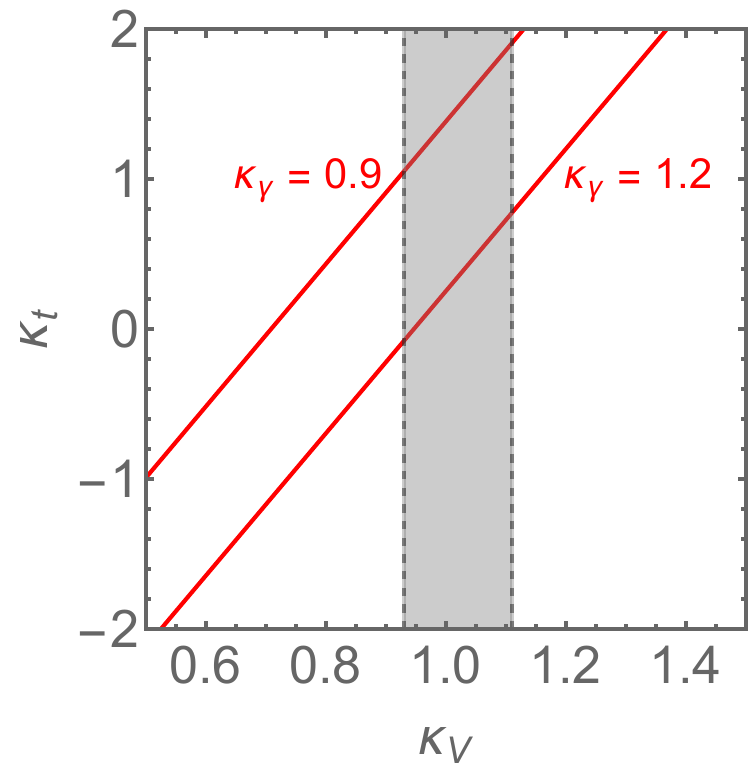}
\caption{Contour of 2$\sigma$ allowed region of $\kappa_\gamma$ as a function of
$\kappa_V^{}$ and $\kappa_t\,(=\kappa_u)$. Shaded regions are allowed at 95\% CL by the current measurement of $\kappa_V^{}$ at the LHC. }
\label{fig:kappa-top}
\end{center}
\end{figure}

We discuss whether WS scenarios are consistent with the measurement of the Higgs signal strength\footnote{Throughout this subsection, 
we take $M_{H^ \pm } = 200$ GeV, and $\Lambda_3$ is fixed such that $\Lambda_2=4\pi$ is realized for the calculation of the charged Higgs boson loop in the $h\to\gamma\gamma$ decay. 
Such an effect on the $\chi^2$ analysis is not significant.}.
First of all, we consider the WS for the top Yukawa coupling. 
In this case,  the decay rate of the $h \to \gamma\gamma$ process is significantly enhanced, because  
the destructive interference between the $W$ boson and top quark loops in the SM alters constructive. 
In Fig.~\ref{fig:kappa-top}, we show the contour plot of $\kappa_\gamma \equiv \sqrt{\Gamma(h \to \gamma\gamma)/\Gamma(h_{\rm SM} \to \gamma\gamma)}$ as a function 
of $\kappa_V^{}$ and $\kappa_t (=\kappa_u)$. 
It is evident that $\kappa_V^{}$ needs to deviate substantially
from unity to accommodate the WS top Yukawa, i.e., $\kappa_V^{}\lesssim 0.8$ for $\kappa_t = -1$ in order to satisfy 
the constraint $0.9 < \kappa_\gamma < 1.2$ at 95\% CL provided by the LHC~\cite{ATLAS:2020qdt}. 
Such a large deviation in $\kappa_V^{}$ has already been excluded by the LHC data~\cite{ATLAS:2020qdt,CMS:2020gsy}, so that the scenario with the WS top Yukawa is now highly disfavored. 
We thus consider the WS scenario in Yukawa couplings for down-type quarks and/or charged leptons in the following discussion. 

\begin{figure}[t]
\begin{center}
\includegraphics[width=7cm]{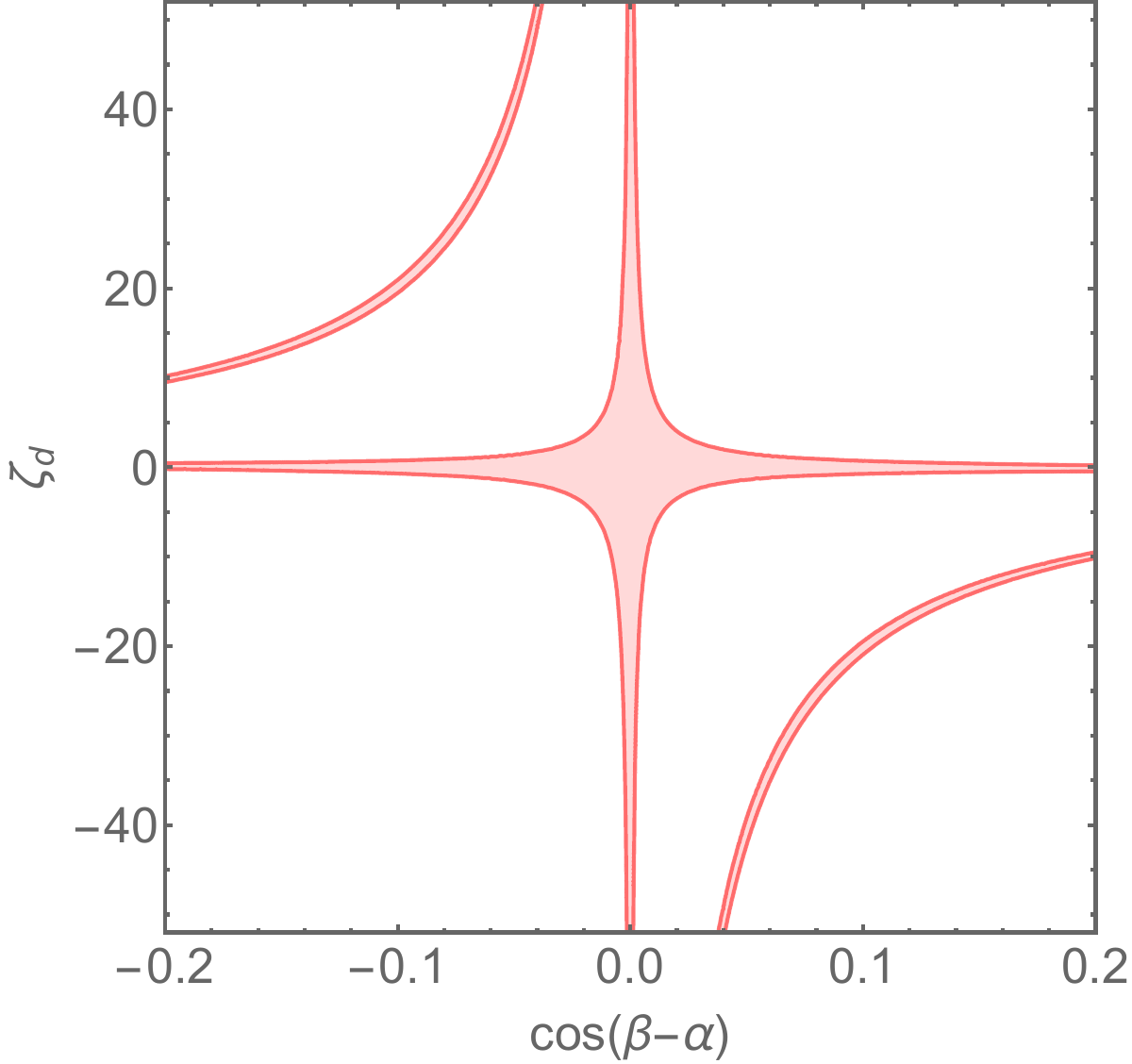}~~
 \includegraphics[width=7cm]{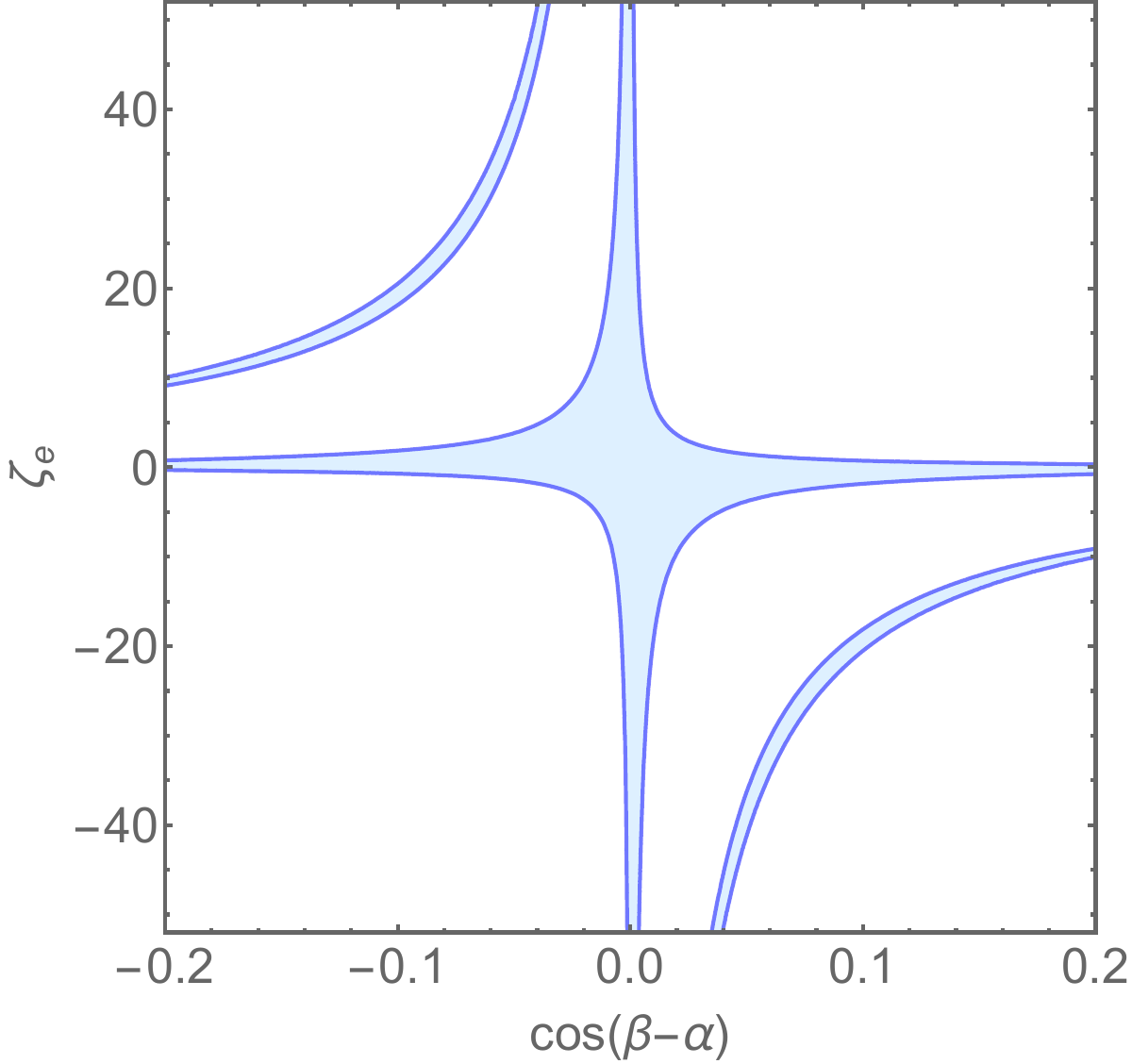}
\caption{Shaded regions are the parameter space allowed at 2$\sigma$ level from the measurements of the Higgs signal strength at the LHC. 
We show the allowed region on the $\cba$-$\zeta_d$ (left panel) and $\cba$-$\zeta_e$ (right panel) plane.  We fixed $M_{H^ \pm } = 200$ GeV, 
$\Lambda_3$ is fixed such that $\Lambda_2=4\pi$, and all remaining parameters were scanned. }
\label{fig:h-chi-sq}
\end{center}
\end{figure}

We apply a $\chi^2$ analysis to constrain the parameter space based on the data shown in Appendix~\ref{app:higgs-data}.
We scan the relevant parameters $\sba$, $\zeta_u$, $\zeta_d$ and $\zeta_e$, and calculate $\Delta \chi^2$ for each scanned point. 
We find that the minimum of $\chi^2$ is given at $\kappa_V=1.02,~\kappa_u = 1.01,~\kappa_d = 1.03$ and $\kappa_e = 0.96$.
The parameter space allowed at $2\sigma$ level is shown in Fig.~\ref{fig:h-chi-sq}. We see that two separate
regions satisfy the Higgs data. The central part corresponds to the RS scenario, while
the bands in the top-left and bottom-right correspond to the WS scenario, where $\zeta_{d/e}\cba\simeq -2$ is satisfied. 
We note that in the $\ztwo$ symmetric 2HDMs, the WS scenario is possible only for $\cos(\beta-\alpha)>0$ because of the structure of the Yukawa coupling given in Eq.~(\ref{eq:types}). 

\begin{figure}[t]
\begin{center}
\includegraphics[width=7.5cm,height=6cm]{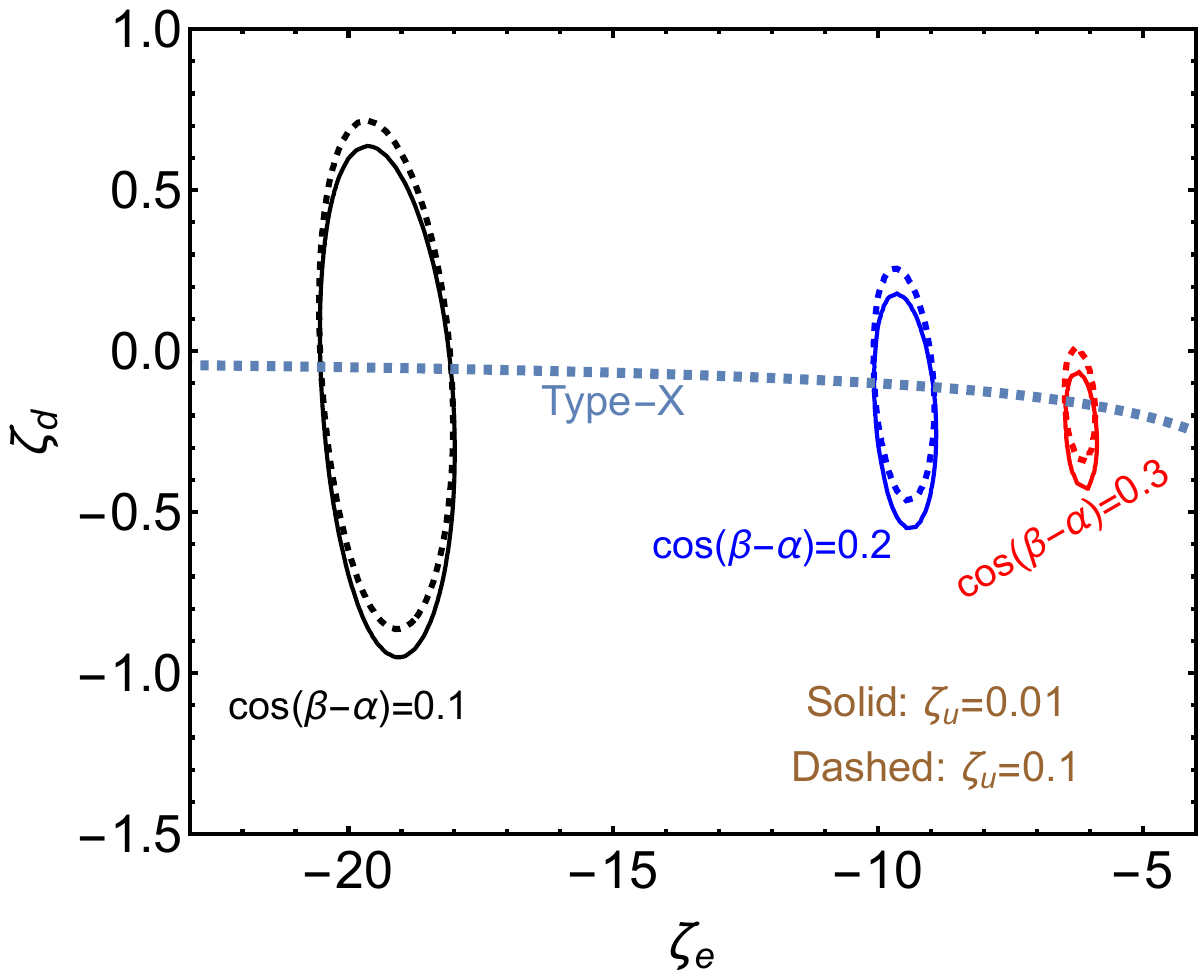}~~
\includegraphics[width=7.5cm,height=6cm]{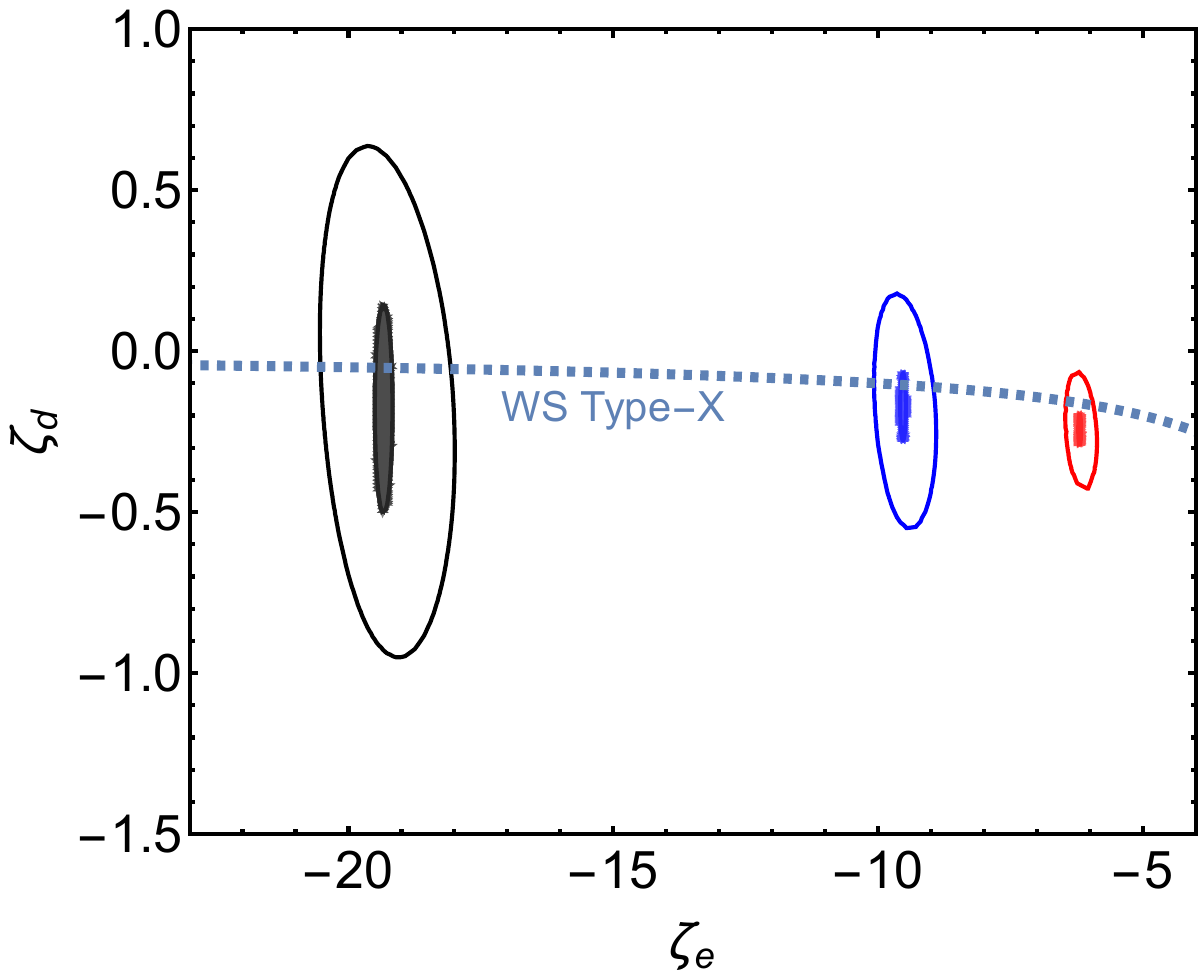}
\caption{Regions inside the ellipses are allowed by the measurement of the Higgs signal strength at $2\sigma$ level on 
the $\zeta_d$-$\zeta_e$ plane for fixed values of $\cba$ and $\zeta_u = 0.1$. 
 The filled ellipses in the right panel shows projected allowed region at the 
HL-LHC. The projected uncertainties are taken from~\cite{Dainese:2019rgk}. 
The dashed curve shows the case realized in the Type-II (Type-X) 2HDM in the left (right) panel. 
 }
\label{fig:yb-ytau}
\end{center}
\end{figure}

Fig.~\ref{fig:yb-ytau} shows the parameter space allowed by the constraints from the Higgs signal strength within $2\sigma$ 
on the $\zeta_e$-$\zeta_{d}$ plane for a fixed value of $\cba$ and $\zeta_u$. 
In this figure, the dotted curve shows the case in the Type-II (left) and Type-X (right) 2HDM. 
We see that small deviations from these 2HDMs shown by the ellipses are allowed, and these are realized in the A2HDM. 

It is worth mentioning here that the WS scenario typically requires large $\tan\beta$ in the 2HDMs with the $\ztwo$ symmetry, which 
gives $\kappa_d = \kappa_e = -1$ ($\kappa_e = -1$) and $\kappa_u \simeq \kappa_V^{}$ ($\kappa_u \simeq \kappa_d \simeq \kappa_V^{}$) in the WS Type-II (WS Type-X) 2HDM. 
On the other hand, in the A2HDM, 
even if we impose the WS condition for, e.g., $\kappa_e$, the $\kappa_u$, $\kappa_d$ and $\kappa_V^{}$ parameters 
can differently deviate from unity because these are controlled by $\zeta_u$, $\zeta_d$ and $\sin(\beta-\alpha)$, respectively. 
Therefore, by measuring the pattern of deviations in the $h$ couplings from the SM predictions, 
we can distinguish the WS scenario in the A2HDM from that in the $\ztwo$ symmetric 2HDMs, see also Ref.~\cite{Kanemura:2014bqa,Kanemura:2015mxa} for the 
model discrimination by ``fingerprinting'' the $h$ couplings.

\subsection{Theoretical Constraints}\label{sec:theoretical_constraints}

The quartic couplings are constrained by the requirement of vacuum stability of the scalar potential and perturbative unitarity of the $S$-matrix. 

The condition of the vacuum stability, i.e., bounded from below in any direction with large field values, is given by~\cite{Deshpande:1977rw,Nie:1998yn,Kanemura:1999xf},
 \bea
 &&\Lambda_1 \geq 0,\quad \Lambda_2 \geq 0,\quad \sqrt{\Lambda_1 \Lambda_2}+\Lambda_3 \geq
0,\quad \sqrt{\Lambda_1 \Lambda_2}+\Lambda_3+\Lambda_4-|\Lambda_5| \geq 0 \nonumber\\
 &&\dfrac{1}{2}(\Lambda_1 +
\Lambda_2)+\Lambda_3+\Lambda_4+\Lambda_5-2|\Lambda_6+\Lambda_7| \geq 0.
 \eea

\begin{figure}[t]
\begin{center}
\includegraphics[width=7.5cm,height=6cm]{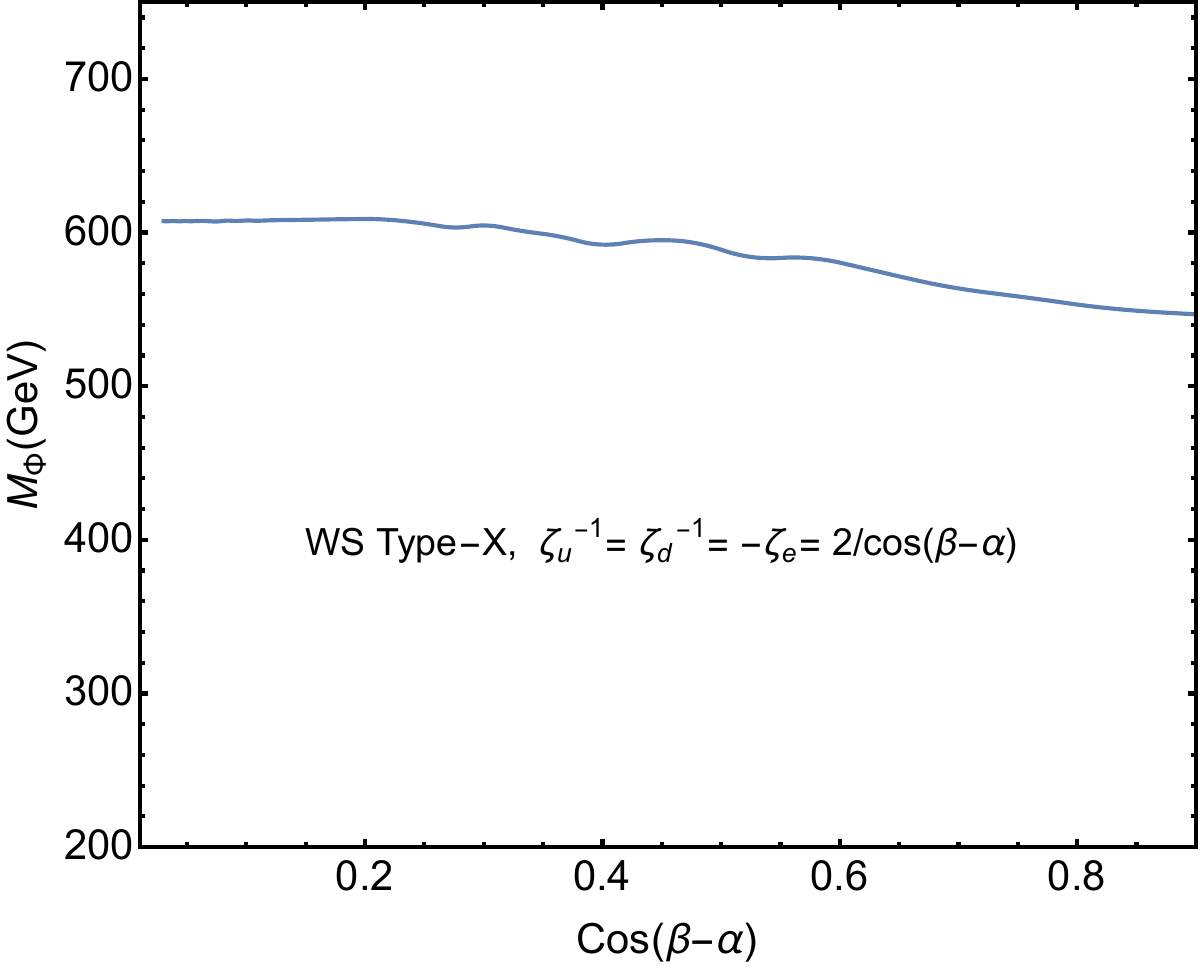}~~
\includegraphics[width=7.5cm,height=6cm]{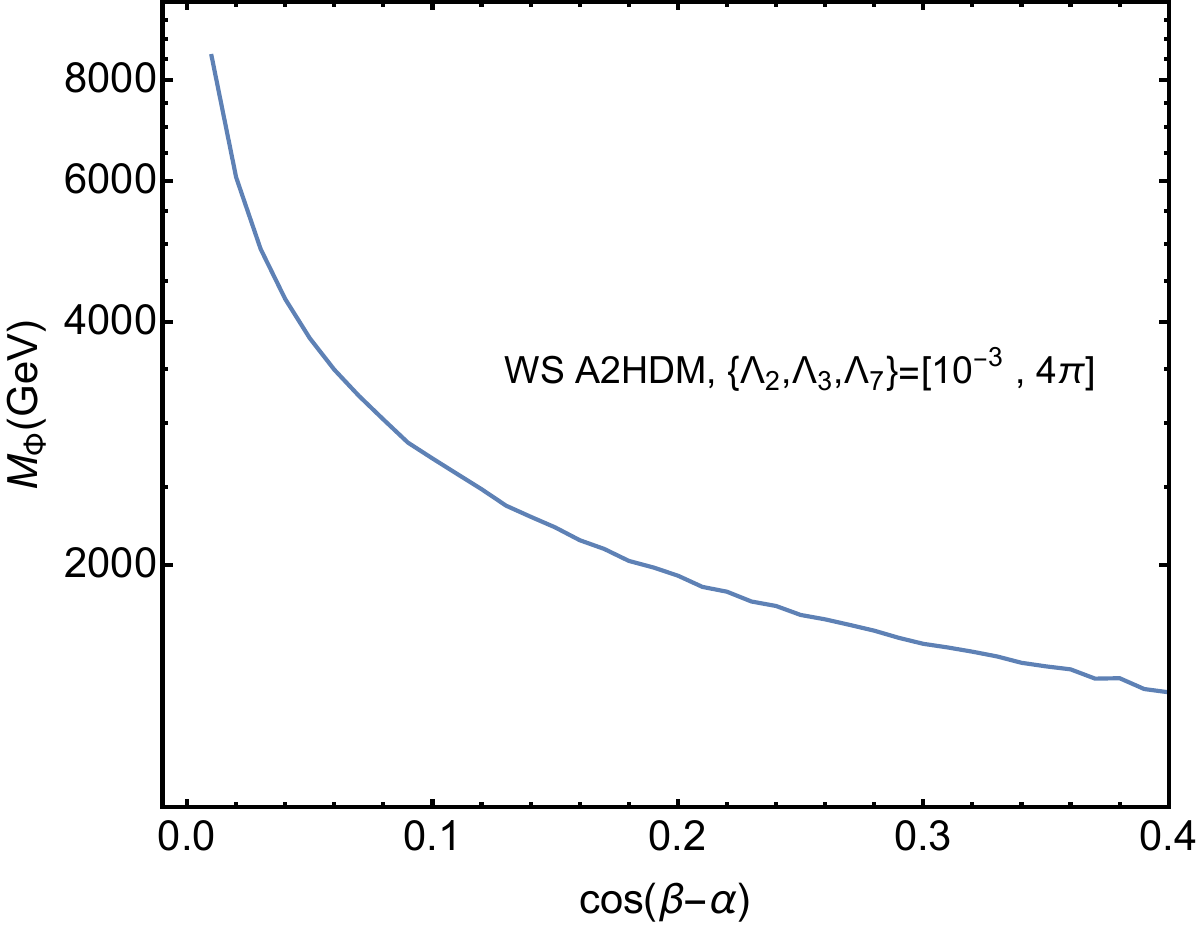}
\caption{Upper limit on the mass of the additional Higgs bosons $M_\Phi = M_{H^\pm} (=M_A = M_H)$ from the 
perturbative unitarity and the vacuum stability as a function of $\cba$ in the 2HDMs with the softly-broken $\mathbb{Z}_2$ symmetry (left) and the A2HDM (right).}
\label{fig:unit-z2}
\end{center}
\end{figure}

For the perturbative unitarity, we consider 2-body to 2-body elastic scatterings for bosonic states in the high-energy limit. 
The analytic expressions for the eigenvalues of the $s$-wave amplitude matrix have been 
found in the $\mathbb{Z}_2$ symmetric case in Refs.~\cite{Kanemura:1993hm,Akeroyd:2000wc,Horejsi:2005da}. 
In the general 2HDM, the $s$-wave amplitude matrix is expressed by a block diagonal form with maximally $4\times 4$ block matrices~\cite{Ginzburg:2005dt,Kanemura:2015ska} which 
can be numerically diagonalized. 
The limit on the masses due to vacuum stability, unitarity and
perturbativity ($|\Lambda_i|< 4\pi$) is shown in Fig.~\ref{fig:unit-z2}. 
For simplicity, we here assume degenerate masses of the additional Higgs bosons, i.e., 
$M_\Phi \equiv M_{H^\pm} (= M_A = M_H)$. All the other relevant parameters are scanned with a sufficiently large range to maximize 
the upper limit on $M_\Phi$. 
The left panel shows the result in the $\mathbb{Z}_2$ symmetric case with the exact WS condition, i.e., $\tan\beta = 2/\cos(\beta-\alpha)$, where 
large values of $\tan\beta$ are required for the case with the nearly Higgs alignment $\cos(\beta-\alpha)\simeq 0$.
In such a case with $\tan\beta \gg 1$ and the nearly Higgs alignment $\cos(\beta-\alpha) \sim 0$, the $\Lambda_2$ parameter can be 
expressed by using Eq.~(\ref{eq:lam237}) as 
\begin{align}
\Lambda_2 &\simeq \frac{\tan^2\beta}{4v^2}\Big[(M_h^2 - M_H^2)\sin2(\beta-\alpha)\tan\beta \notag\\
& + M_H^2(1 - 3\cos2(\beta-\alpha)) + M_h^2(1 + 3\cos2(\beta-\alpha)) -4 M_{H^\pm}^2 + 2 v^2 \Lambda_3 \Big]. \label{eq:lam2-limit}
\end{align}
We see that in order to satisfy the unitarity constraint, a cancellation among the terms in the square bracket is required, and it turns out to be 
a severe upper bound on $M_\Phi$. 
In fact, we numerically find that $M_\Phi$ has to be smaller than about 600 GeV in the WS scenario. 
On the contrary, in the A2HDM (right panel), $\Lambda_2$ is the independent free parameter, so the large cancellation 
mentioned above is not needed. 
As a result, the limit on $M_\Phi$ is much more relaxed as compared with the $\ztwo$ symmetric case.
However, very large $M_\Phi$ is restricted as the couplings become too large as long as $\cos(\beta-\alpha) \neq 0$, see Eq.~(\ref{eq:lam6}). 
\subsection{Flavor Constraints}

 \begin{figure}
 \centering
 \includegraphics[width=100mm]{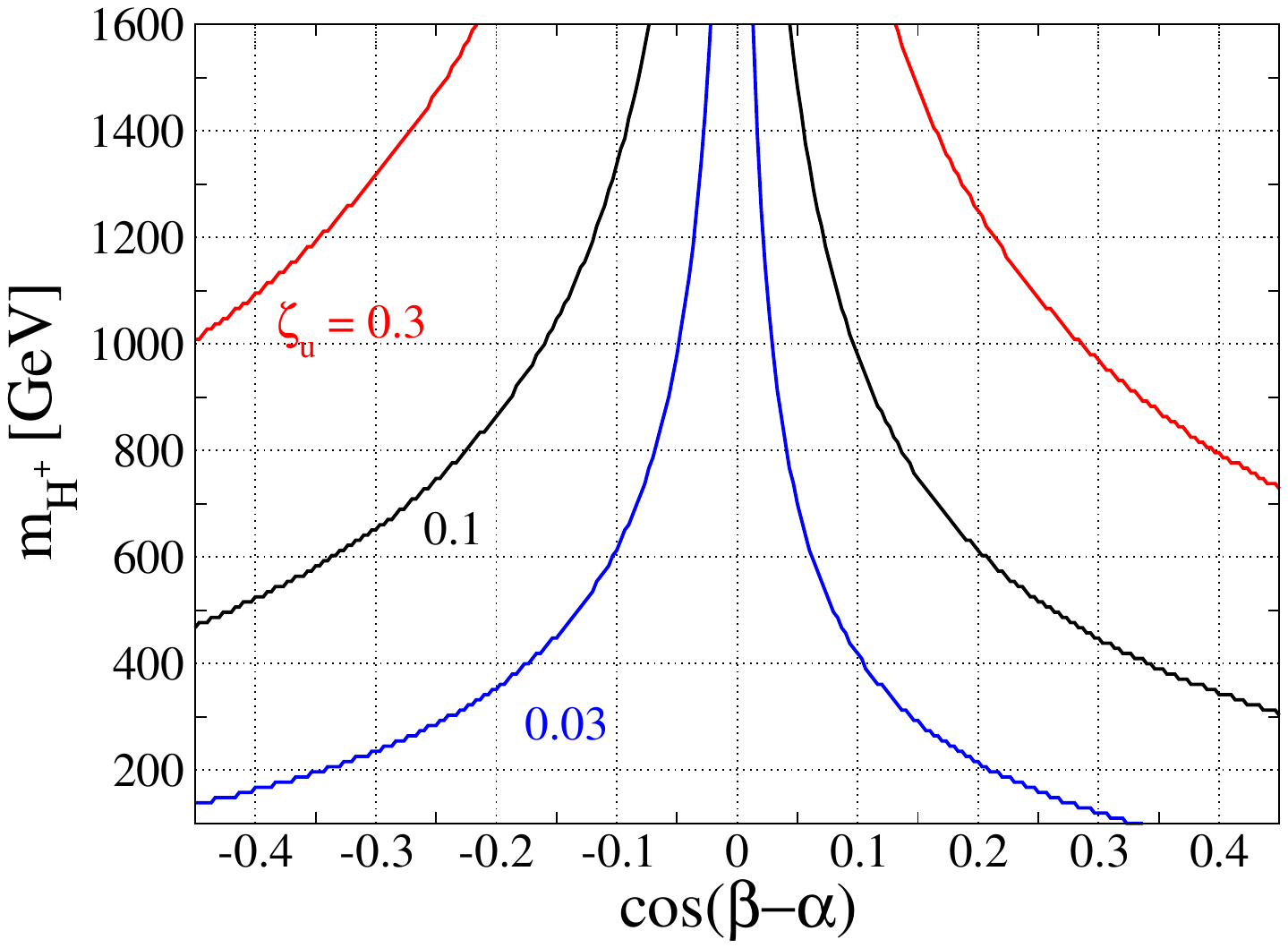}
 \caption{Constraint from $B\to X_s\gamma$ for the case with the WS limit for
 down-type quarks. Regions below each curve is excluded at 95\% CL.}
 \label{fig:bsg}
 \end{figure}

We discuss constraints from flavor experiments.
It is well known that the charged Higgs boson mediates in the $B \to X_s\gamma$
decay process at the one-loop level, by which its mass $M_{H^ \pm }$ and parameters related to
quark Yukawa couplings $\zeta_u$ and $\zeta_d$ are constrained. In the WS limit
for down-type quarks, $\zeta_d$ is determined by fixing $\cos(\beta-\alpha)$, so that the
branching ratio of the $B \to X_s\gamma$ decay is given in terms of $M_{H^\pm}$, $\zeta_u$ and $\cos(\beta-\alpha)$.

The current world average for the measurement of the branching ratio of $B \to X_s\gamma$
is given by~\cite{HFLAV:2019otj}
\begin{align}
  {\cal B}(B \to X_s\gamma) = (3.32 \pm   0.15)\times 10^{-4}.
\end{align}
We implement QCD and QED corrections at next-to-leading order (NLO) calculated in
Refs.~\cite{Borzumati:1998tg} and \cite{Kagan:1998ym}.
We apply the same SM inputs given in Ref.~\cite{Kanemura:2021dez} to the numerical
analysis, by which we obtain the SM prediction to be ${\cal B}(B \to X_s \gamma) =
3.24\times 10^{-4}$. In Fig.~\ref{fig:bsg}, we show the lower limit on the value of
$M_{H^ \pm }$ as a function of $\cos(\beta-\alpha)$ for fixed values of $\zeta_u$ to be
0.3 (red), 0.1 (black) and 0.03 (blue) from the data of $B \to X_s\gamma$ decay.
We see that stronger bounds on $M_{H^ \pm }$ are obtained for smaller values of
$|\cos(\beta-\alpha)|$ corresponding to larger $|\zeta_d|$ due to the WS condition
and/or larger values of $\zeta_u$, because the decay amplitude is enhanced by $\zeta_u^2$
and/or $\zeta_u\zeta_d$ terms. It is also seen that the case with $\cos(\beta-\alpha)
< 0~(\zeta_d > 0)$ gives a stronger bound on $M_{H^ \pm }$ as compared to that with
$\cos(\beta-\alpha) > 0~(\zeta_d < 0)$ for a fixed value of $\zeta_u$. This is because in
the case with $\zeta_d > 0$ the new physics contribution, proportional to
$\zeta_u\zeta_d$, gives a destructive interference with the SM contribution, so it reduces
the branching ratio. As we show above, the SM prediction is given below the central value,
stronger limits are provided for $\cos(\beta-\alpha)<0$. We note that in the Type-II case
$\zeta_u$ is fixed to be $-1/\zeta_d (= \cot\beta)$, so that the new physics contribution
almost do not depend on $\tan\beta$ for $\tan\beta \gtrsim 2$, and the current lower limit
on $M_{H^ \pm }$ has been taken to be around 800 GeV based on the calculation with
next-to-NLO QCD~\cite{Misiak:2017bgg}. We also note that the other flavor constraints,
such as $B_s \to \mu^+\mu^-$ are typically much weaker than that from $B \to X_s\gamma$ as
long as we do not take extremely large value of $|\zeta_e|$, see e.g.,
Ref.~\cite{Kanemura:2021dez}.


\section{Decays of Additional Higgs Bosons in RS \& WS Scenarios}\label{sec:comparison}

We compare decay branching ratios of the additional Higgs bosons in the WS and RS scenarios.
Since the Type-II scenario has already been strongly constrained by LHC data and flavor observables~\cite{Su:2019ibd}, 
we here focus on the phenomenology in the Type-X 2HDM and the A2HDM with $|\zeta_e|\gg |\zeta_{u/d}|$. 
As we discussed in Sec.~\ref{sec:constraints}, in the WS scenario 
larger values of $|\cba|$ are required for $|\zeta_e| \gg 1$ as compared with the RS scenario, 
the phenomenology will be different from the RS region. 
Now, we present the difference in decays of the additional Higgs bosons in the RS and
WS scenarios. For our phenomenological analysis, we assume that the additional Higgs bosons are
degenerate in mass, i.e., $M_\Phi \equiv M_{H^\pm} (=M_A=M_H)$.

For the $H \to hh$ mode, the decay rate depends on the scalar trilinear coupling $\lambda_{Hhh}$ defined as the coefficient of the $Hhh$ vertex in the Lagrangian, which is given by 
 \bea
 \lambda_{Hhh} = -\dfrac{\cos(\beta-\alpha)}{2v}&&
\bigg\{M_H^2 - 4 M_{H^ \pm}^2 + \cos^2(\beta-\alpha) \left(6 M_{H^ \pm }^2 - 2M_h^2 - M_H^2\right)\nonumber\\
&& + {v}^2\left[\frac{3}{2} \sin 2(\beta-\alpha)
\Lambda_7 + (2-3\cos^2(\beta-\alpha)) \Lambda_3\right],\nonumber \\
\simeq  -\dfrac{\cos(\beta-\alpha)}{2v}&&  \bigg\{M_H^2 - 4 M_{H^ \pm}^2 + {v}^2\left[\frac{3}{2} \sin 2(\beta-\alpha)
\Lambda_7 + 2 \Lambda_3\right]
\bigg\}, \label{eq:Hhh}
 \eea
where the last expression is valid for $\sin(\beta-\alpha)\simeq 1$. 
Thus, it depends on the additional parameters $\Lambda_3$ and $\Lambda_7$, which do not appear in the other decay modes 
considered in this section. In the A2HDM, we take $\Lambda_3 = \Lambda_7 = 0$ for a while, and later we show how the branching 
ratio of $H \to hh$ is changed depending on the $\Lambda_3$ and $\Lambda_7$ parameters. 
In the Type-X 2HDM, 
$\Lambda_7$ is determined by Eq.~(\ref{eq:lam7}), while $\Lambda_3$ is in principle free parameter, 
but its value is substantially determined by the unitarity bound. 
This can be seen from Eq.~(\ref{eq:lam2-limit}), i.e., $\Lambda_3$ is determined in such a way that terms in the square 
bracket of Eq.~(\ref{eq:lam2-limit}) vanish in order to maintain $\Lambda_2$ to be order 1. 
We apply the above treatment for $\Lambda_3$ and $\Lambda_7$ to the following calculation.

\begin{figure}[t]
\begin{center}
\includegraphics[width=8cm]{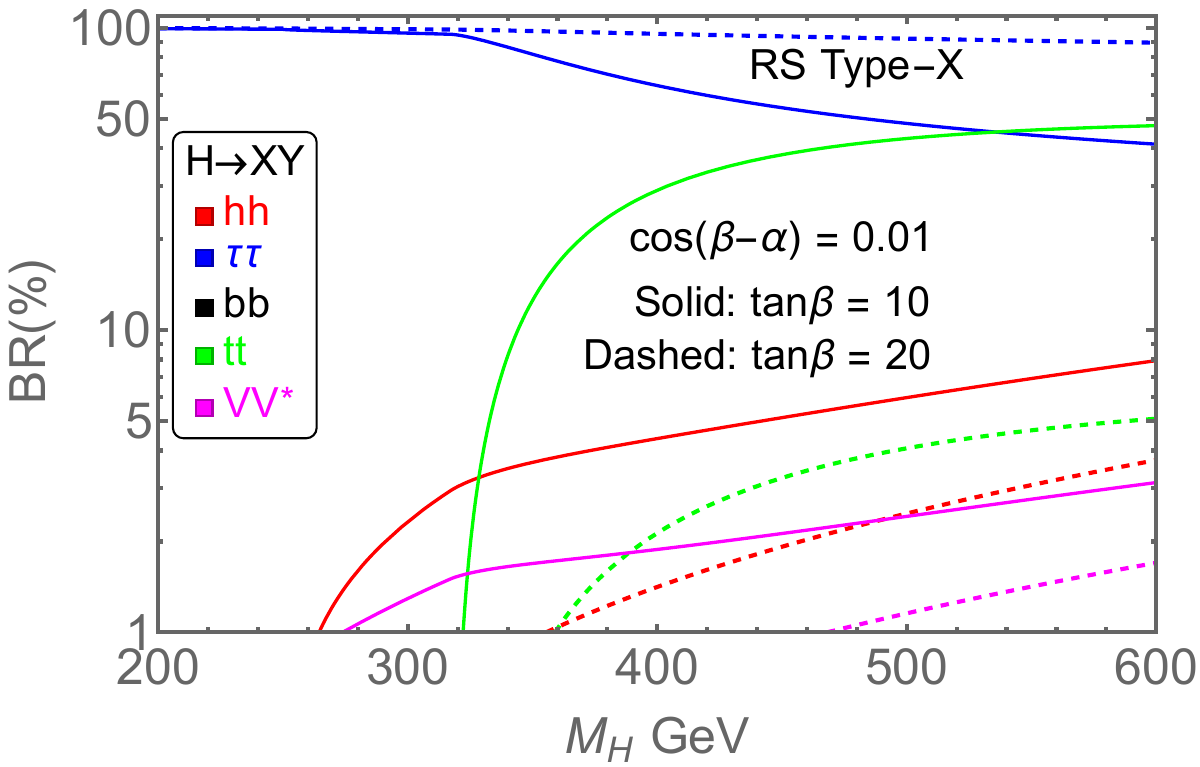}~~
\includegraphics[width=8cm]{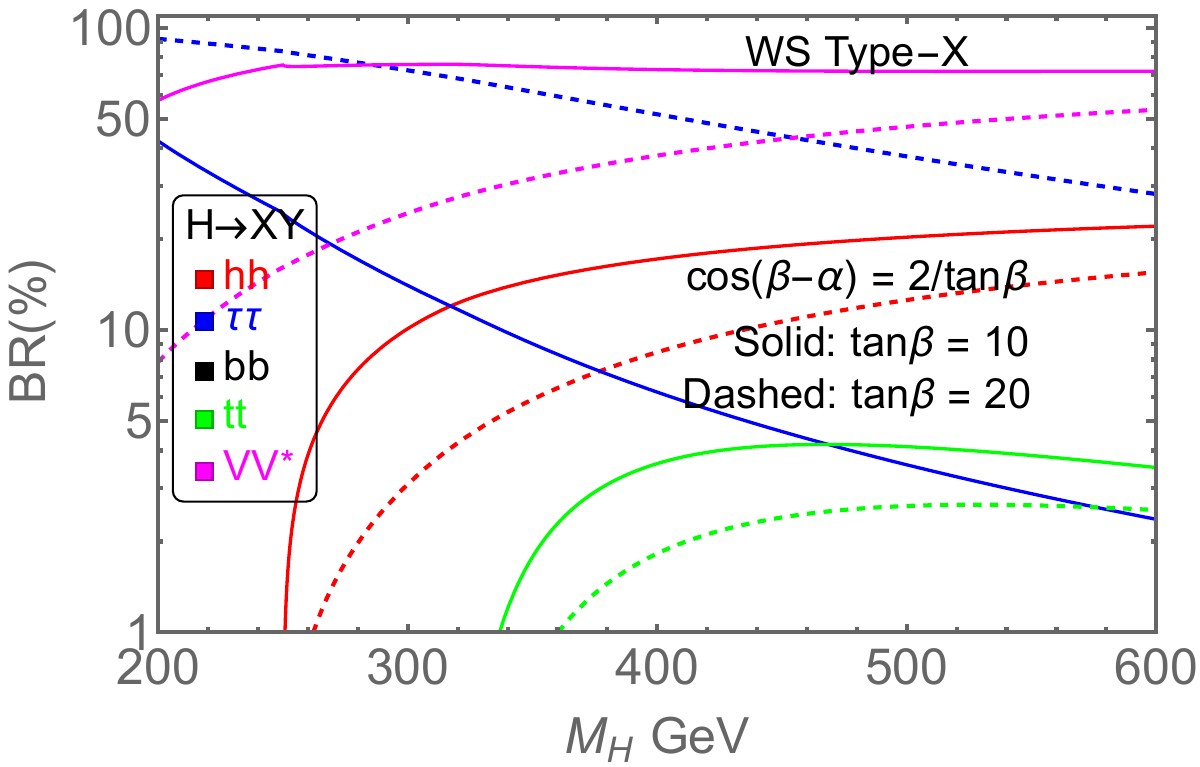}
\includegraphics[width=8cm]{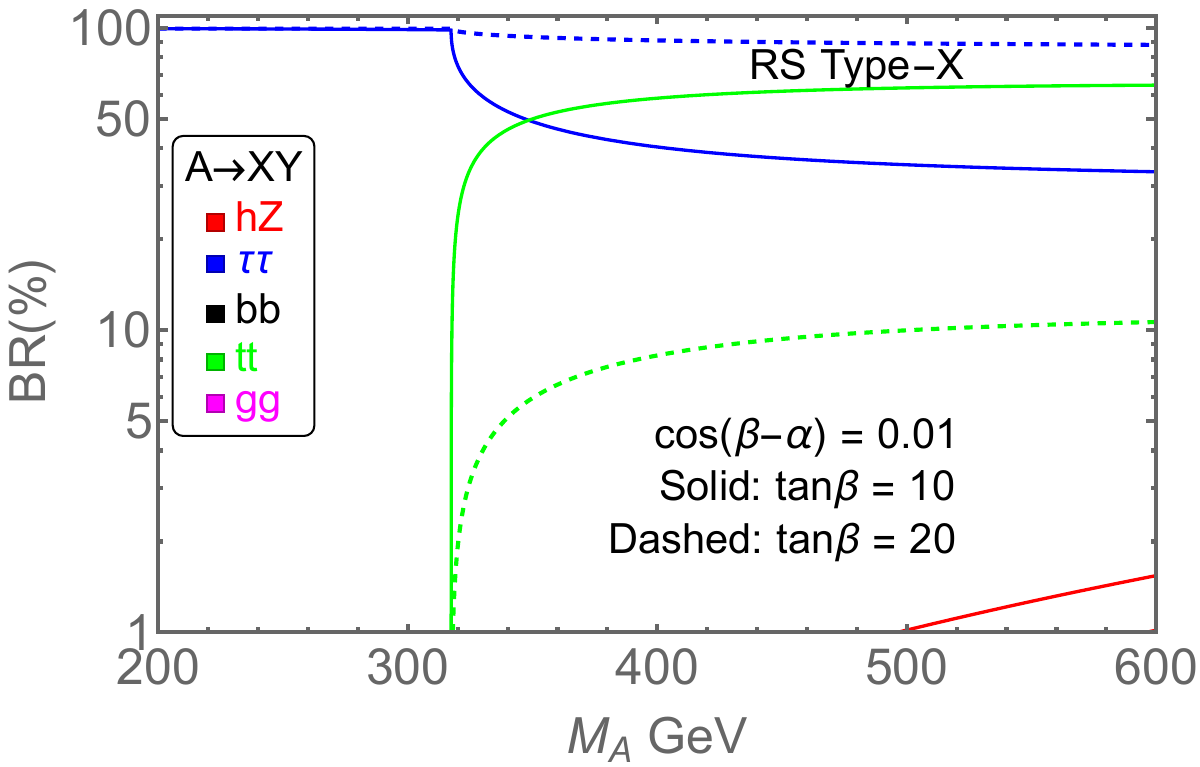}~~
\includegraphics[width=8cm]{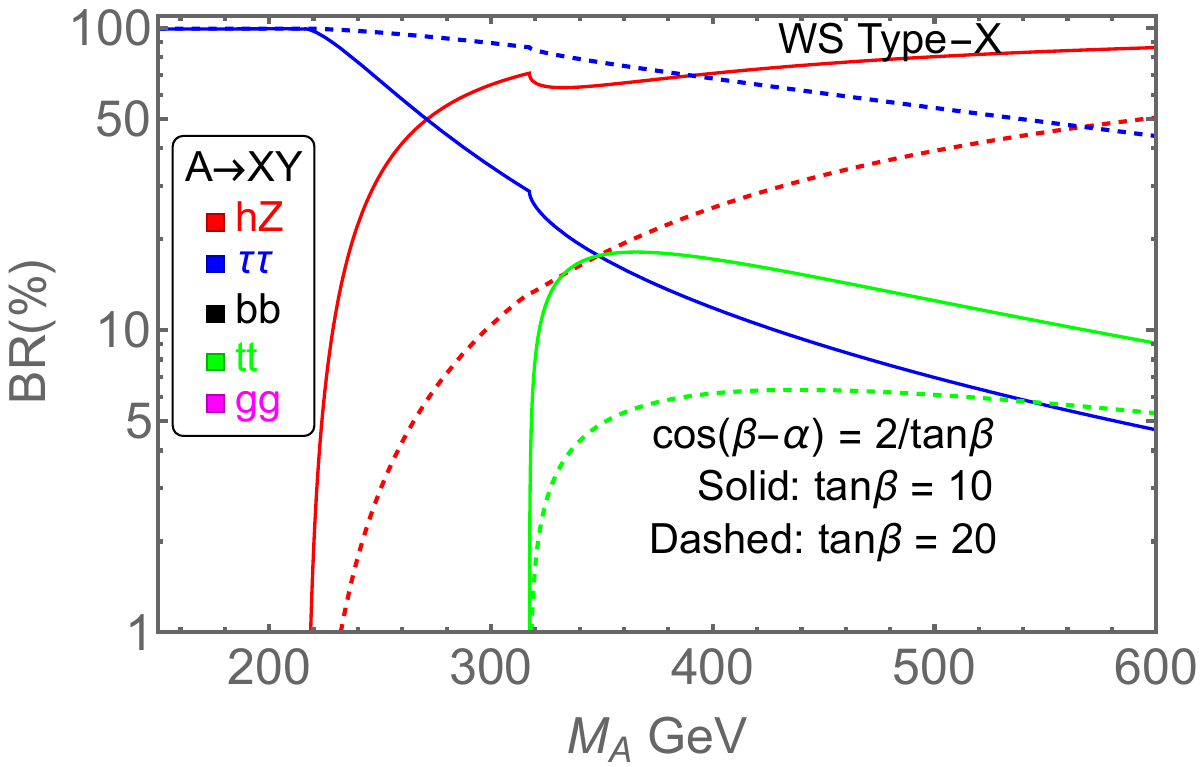}
\caption{Comparison of branching ratios for the additional Higgs bosons ($H$: upper panels and $A$: lower panels)
in the RS Type-X 2HDM with $\cos(\beta-\alpha) = 0.01$ (left panels) and the WS Type-X 2HDM with $\cos(\beta-\alpha) = 2/\tan\beta$ (right panels). 
The solid and dashed curves respectively show the case with $\tb=10$ and 20.}
\label{fig:compare-X}
\end{center}
\end{figure}

In Fig.~\ref{fig:compare-X}, we show the branching ratios of $H$ (upper panels) and $A$ (lower panels) in the Type-X 2HDM with 
$\tan\beta = 10$ (solid curves) and 20 (dashed curves). 
The left and right panel respectively focuses on the RS and WS scenario, where the value of $\cos(\beta-\alpha)$ is fixed to be 
0.01 (almost the upper limit from the Higgs signal strength) and $2/\tan\beta$ which gives $\kappa_\tau = -1$ and 
$\kappa_b\simeq\kappa_t \simeq 1$.
As expected, for the RS case, the $H \to \tau\tau$ mode is dominant, and $t\bar{t}$ can also be dominant for $\tan\beta = 10$ 
and $M_{H}>2m_t$. On the other hand, in the WS case, $H \to VV$ ($V=W^\pm$, $Z$) can be dominant instead of $\tau\tau$, 
particularly for the case with larger $M_\Phi$. In addition, the $H\to hh$ decay can also be substantial. 
Similar behavior can be seen in the decay of $A$, but the $A \to Zh$ can be dominant in the WS case as the $VV$ and $hh$
 modes are forbidden for $A$. Therefore, it is clear that the decays into bosonic states become more important than the fermionic 
 one in the WS scenario, which can be important to distinguish the WS scenario from the RS one. 
\begin{figure}[t]
\begin{center}
\includegraphics[width=8cm]{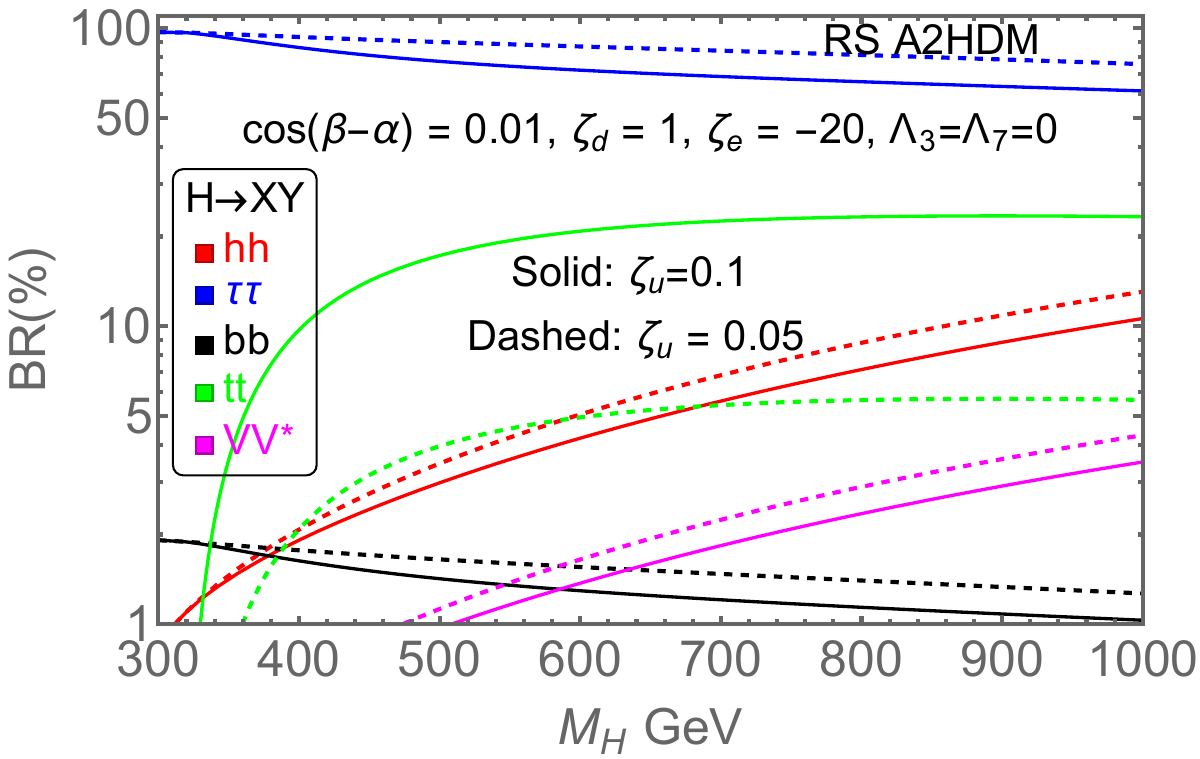}~~
\includegraphics[width=8cm]{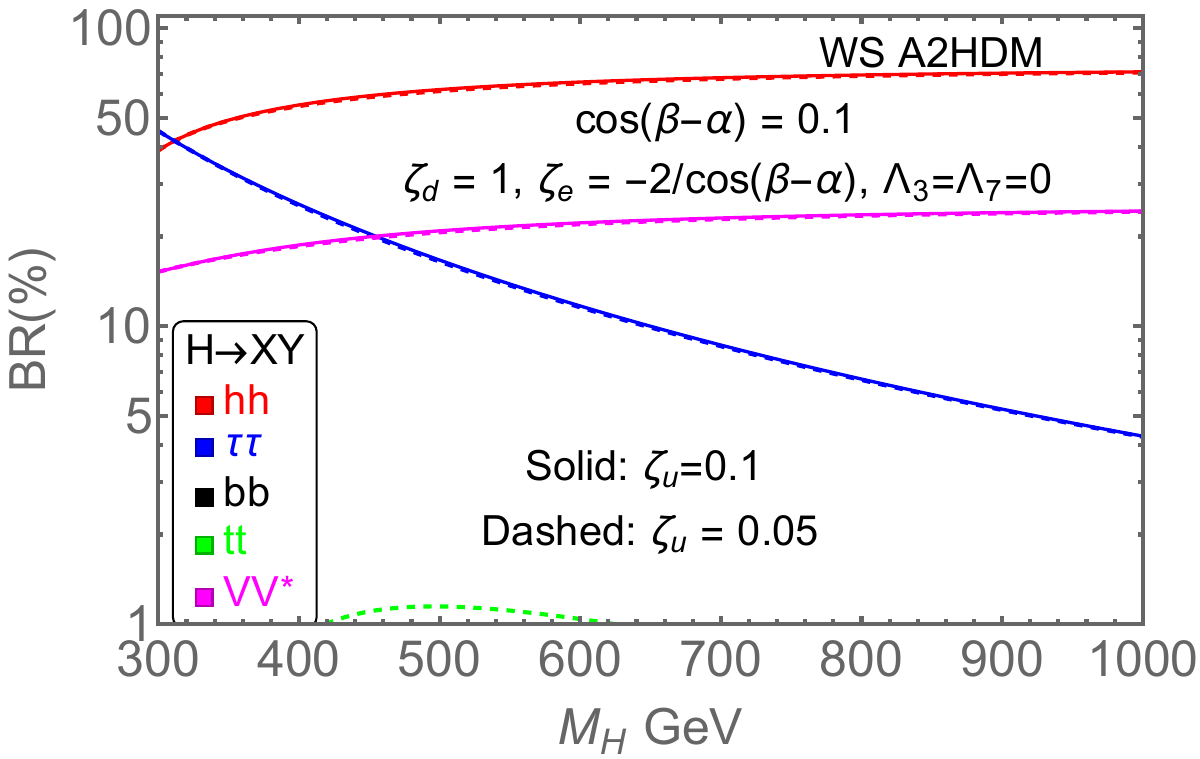}\vspace{0.2cm}
\includegraphics[width=8cm]{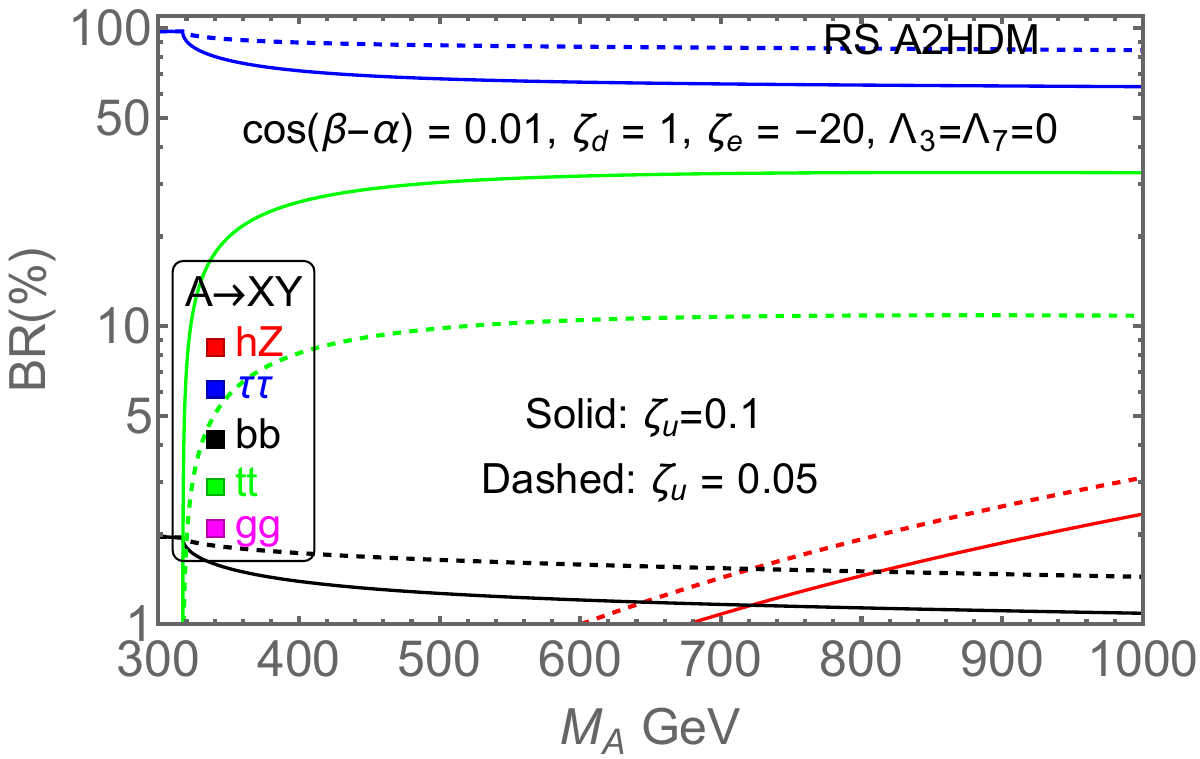}~~
\includegraphics[width=8cm]{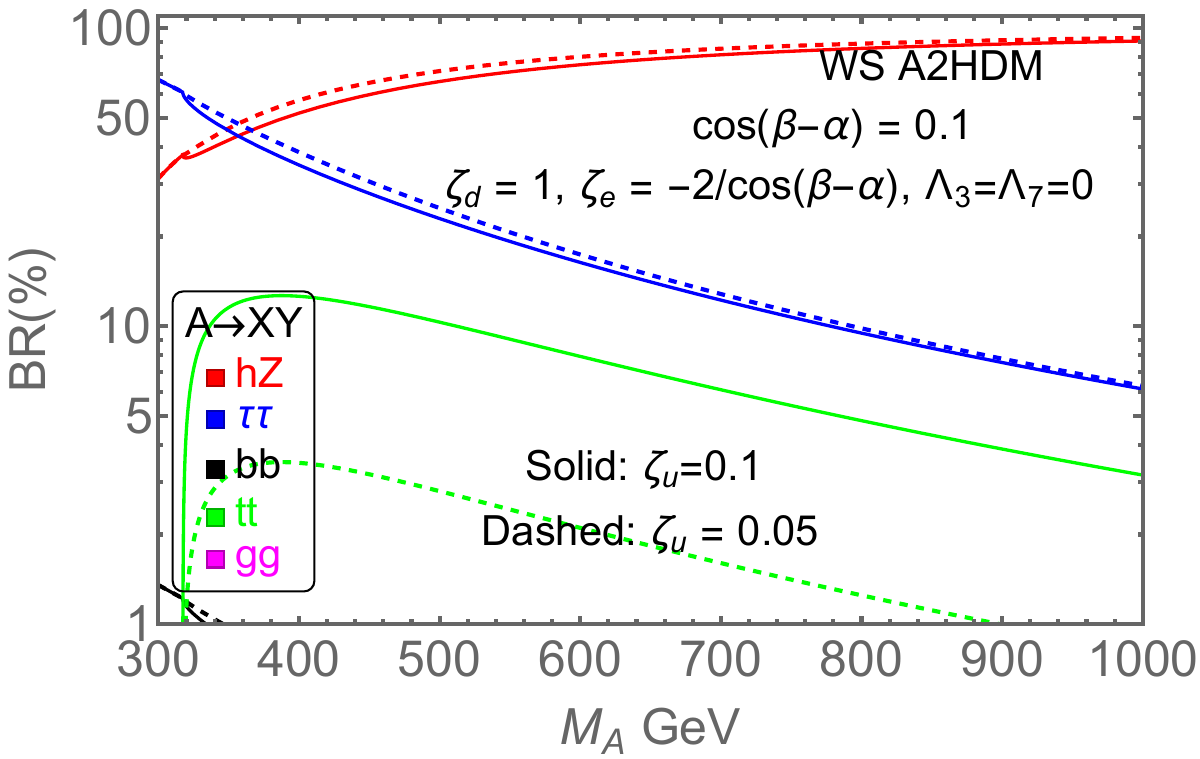}
\caption{Comparison of branching ratios for the additional Higgs bosons ($H$: upper panels and $A$: lower panels)
in the RS A2HDM with $\cos(\beta-\alpha) = 0.01$ (left panels) and the WS A2HDM with $\cos(\beta-\alpha) = 2/\tan\beta$ (right panels). 
We fix $\zeta_d = 1$, $\zeta_e = -20$ and $\Lambda_3 = \Lambda_7 = 0$ for all the panels, while 
The solid and dashed curves respectively show the case with $\zeta_u = 0.1$ and 0.05.  }
\label{fig:compare-A2HDM}
\end{center}
\end{figure}

In Fig.~\ref{fig:compare-A2HDM}, we show the branching ratios for $H$ (upper panels) and $A$ (lower panels) in the A2HDM with $\zeta_e = -20$, 
$\zeta_d = 1$ and $\Lambda_3 = \Lambda_7 = 0$. The solid and dashed curves denote the case with $\zeta_u = 0.1$ and 0.05, respectively. 
Again, the left (right) panel shows the result in the RS (WS) case, where we take $\cba=0.01$ ($\cba = -2/\zeta_e$). 
These values mimic the Type-X Yukawa structure. 
Similar to the Type-X case, the decay modes into bosonic states are more important in the WS case as compared with the RS case.
The difference between the WS scenario in the Type-X 2HDM and the A2HDM is seen in the $H \to hh$ decay, which can be more significant in the 
A2HDM than that in the Type-X 2HDM due to the freedom of the $\Lambda_3$ and $\Lambda_7$ parameters.

\begin{figure}[t]
\begin{center}
\includegraphics[width=6cm]{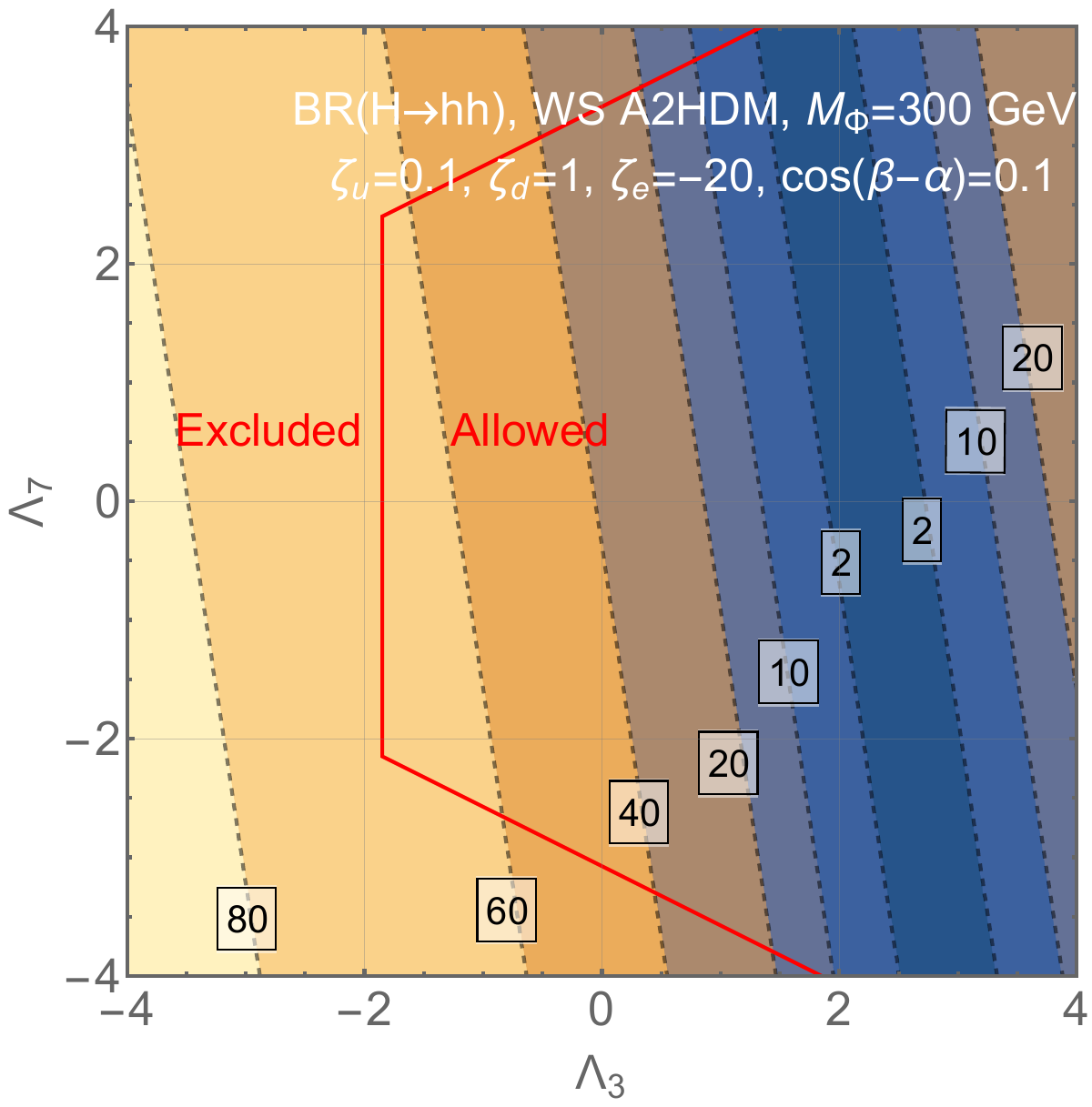}
\hspace{1cm}
\includegraphics[width=6cm]{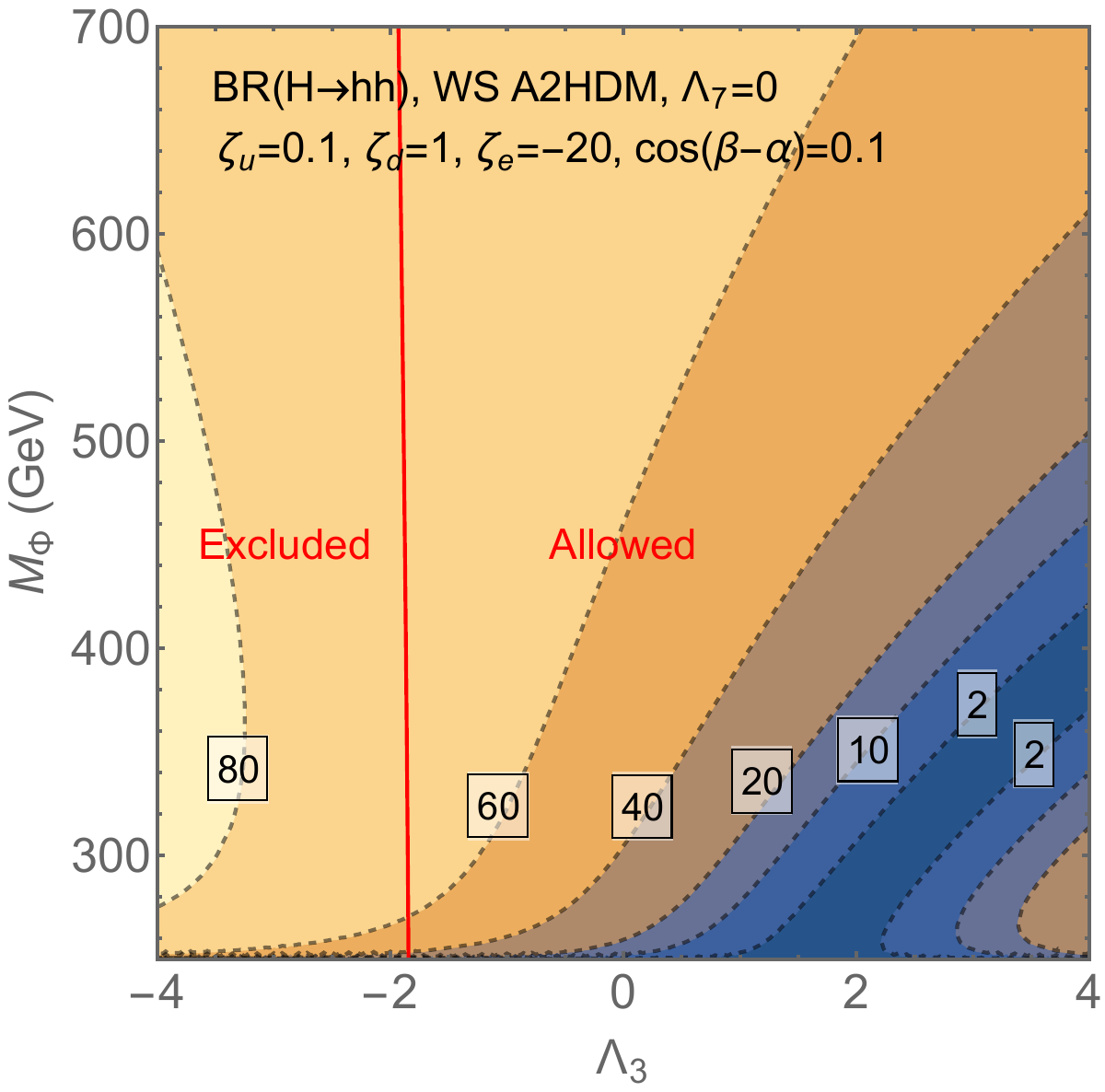}
\caption{Contours of the branching ratio for $H  \to  hh$ in percent on the
$\Lambda_3$-$\Lambda_7$ plane (left)
and the $\Lambda_3$-$M_H$ plane (right).
Constraint from the vacuum stability of the scalar potential is indicated by the solid
curves.}
\label{fig:br-Hhh}
\end{center}
\end{figure}

In Fig.~\ref{fig:br-Hhh}, we show the contours of BR$(H\to hh)$ on the $\Lambda_3$-$\Lambda_7$ plane (left) and the $\Lambda_3$-$M_H$ plane. 
We also depict the region excluded by theoretical constraints as the red-solid line. 
It is seen that a larger branching ratio is obtained for a smaller value of $\Lambda_3$, because it enhances the $\lambda_{Hhh}$ coupling 
as seen in Eq.~(\ref{eq:Hhh}). On the other hand, the dependence of $\Lambda_7$ on BR($H \to hh$) is quite milde due to the suppression 
factor of $\sin2(\beta-\alpha)$, see Eq.~(\ref{eq:Hhh}). The branching ratio can be significantly large values, e.g., about 60\% at 
$(\Lambda_3,\Lambda_7)=(-1,0)$ with $M_H = 300$ GeV. 
We note that the decay rate of $H \to hh$ also strongly depends on the mass spectrum of the additional Higgs bosons as it can be seen 
in Eq.~(\ref{eq:Hhh}). For instance, if we take $M_{H^\pm} - M_{H} = 50~(100)$ GeV with $M_H = 300$ GeV, $M_A = M_{H^\pm}$ and 
$\Lambda_3 = \Lambda_7 = 0$, the branching ratio of $H \to hh$ becomes about $60$\% (70\%), which is allowed by the constraints from 
the perturbative unitarity and vacuum stability as well as the electroweak $T$ parameter due to the custodial symmetry in the Higgs 
potential~\cite{Pomarol:1993mu}.

\section{Limits from Current LHC and HL-LHC}\label{sec:LHC-limit}


We discuss the parameter space giving the WS scenario in the Type-X 2HDM and the A2HDM, which is excluded by direct searches for 
additional Higgs bosons at the LHC Run-II experiment and 
is expected to be explored by the HL-LHC. In the following discussion, we take $\Lambda_3 = \Lambda_7 = 0$, $\zeta_u=0.1$ and $\zeta_d = 1$ 
in the A2HDM, and $\Lambda_3$ is fixed so as to keep $\Lambda_2$ to be order 1 in the Type-X 2HDM as we have done in the previous section. 

The processes which are crucial to constrain the parameter space are the following:
\begin{align}
&pp\to A \to Zh, \quad pp\to A\to \tau\tau,\quad pp\to H \to hh,\quad pp\to H\to ZZ. \label{eq:lhc}
\end{align}
We apply the 95\% CL upper limit on the cross section given at the LHC with the integrated luminosity of 139 fb$^{-1}$ for 
the $Zh$~\cite{ATLAS:2022enb}, $\tau\tau$~\cite{ATLAS:2020zms}, $hh$~\cite{ATLAS:2022hwc} and $ZZ$~\cite{ATLAS:2020tlo} mode. 
Here, all the additional Higgs bosons are assumed to be produced via the gluon fusion process.
We employ \texttt{SusHi}~\cite{Harlander:2012pb,Harlander:2016hcx} in order to estimate the production
cross-section of the gluon fusion at NNLO. 
Since our scenarios are leptophilic, the $b$ quark-associated production ($b\bar bH$ and $b\bar bA$) is negligible.
Other relevant channels are $pp \to H/A \to t\bar t$~\cite{ATLAS:2018rvc} and the charged Higgs boson production from $pp\to tH^\pm$
with the decays $H^\pm \to tb$~\cite{ATLAS:2021upq}, $H^\pm \to \tau^\pm\nu$~\cite{ATLAS:2018gfm} and $H^\pm \to W^\pm h$~\cite{CMS:2022jqc}. 
We, however, find that the limits from these channels are weaker than those from Eq.~(\ref{eq:lhc}). 

\begin{figure}[t]
\begin{center}
\includegraphics[width=8.0cm,angle=00]{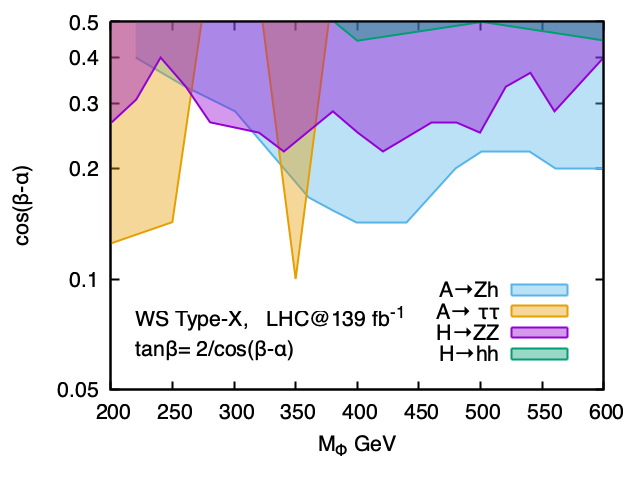}~~
\includegraphics[width=8.0cm,angle=00]{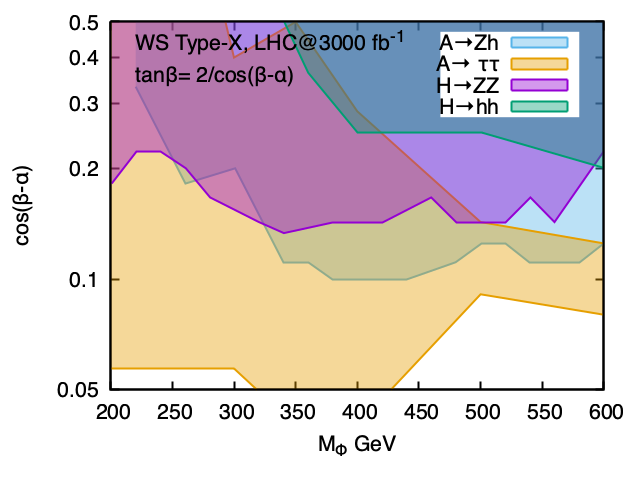}
\caption{Shaded regions on the $M_\Phi$-$\cos(\beta-\alpha)$ plane are excluded at 95\% CL from the direct searches at the current LHC (left) and the HL-LHC (right) in the Type-X 2HDM.}
\label{fig:LHC-limit-X}
\end{center}
\end{figure}

In Fig.~\ref{fig:LHC-limit-X}, we show the parameter space in the Type-X 2HDM, which is excluded by the current LHC data (left) and 
is expected to be excluded by the HL-LHC with the integrated luminosity of 3000~fb$^{-1}$ (right). 
In order to find out the limit expected at the HL-LHC, we calculate the signal cross section required for 2$\sigma$ significance assuming 
the same background cross section and the same 95\% CL upper limit on the event number reported by the Run-II experiments.
In this way, roughly $\sqrt{139/3000}\simeq 0.22$ times smaller signal cross section with respect to that given at the Run-II experiment is excluded at the HL-LHC. 
The colored regions are excluded at 95\% CL
by the processes given in Eq.~(\ref{eq:lhc}). 
As we have seen in Fig.~\ref{fig:unit-z2}, the upper limit on the mass is given to be about 600 GeV from the unitarity constraint, 
so that we show the region up to 600 GeV. 
We see that most of the region with $\cos(\beta-\alpha) \gtrsim 0.2$ is excluded by the combination of the various constraints from the direct searches, while 
that with smaller values of $\cos(\beta-\alpha)$ is allowed, because the production cross section is suppressed by $\cot^2\beta$ which is now fixed by the WS condition, i.e., $\tan\beta \simeq 2/\cos(\beta-\alpha)$. 
At the HL-LHC, the region expected to be excluded is enlarged up to $\cos(\beta - \alpha)\gtrsim 0.1$.

\begin{figure}[t]
\begin{center}
\includegraphics[width=8.0cm,angle=00]{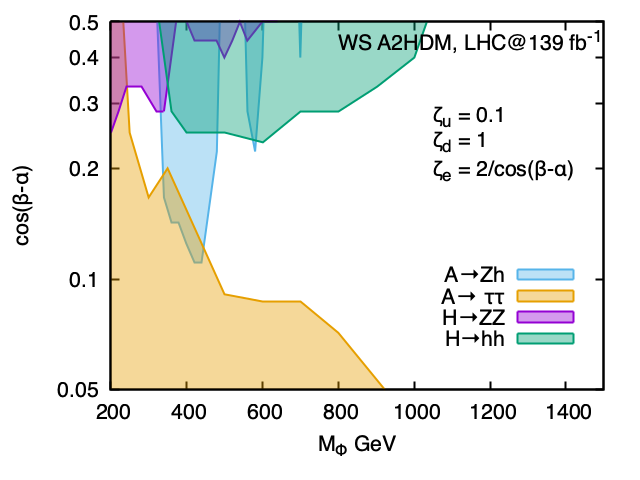}~~
\includegraphics[width=8.0cm,angle=00]{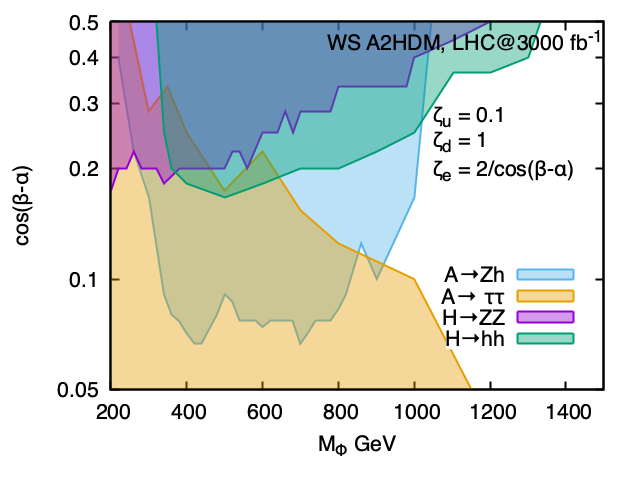}
\caption{Shaded regions on the $M_\Phi$-$\cos(\beta-\alpha)$ plane are excluded at 95\% CL from the direct searches at the current LHC (left) and the HL-LHC (right) in the A2HDM.}
\label{fig:LHC-limit-A}
\end{center}
\end{figure}

In the A2HDM, the LHC limits strongly depend on the value of $\zeta_u$ as the
production of the neutral Higgs bosons via gluon fusion is directly proportional to
$\zeta_u^2$. In Fig.~\ref{fig:LHC-limit-A}, we show the LHC bounds for $\zeta_u=0.1$.
We see that the region with smaller $\cos(\beta-\alpha)$ is excluded mainly by the $A \to \tau\tau$ mode, 
because the decay rate of $A \to \tau\tau$ ($A \to Zh$ and $H \to hh$) is proportional to $\zeta_e^2 \simeq [2/\cos(\beta-\alpha)]^2$. 
On the other hand, the case with larger $\cos(\beta-\alpha)$ is excluded by the $A \to Zh$ and $H \to hh$ modes, because the decay rates of these modes are suppressed by 
$\cos^2(\beta-\alpha)$. 
Thus, these two channels are complementary with each other. 
We note that the limits from the searches for $A$ vanish for $\zeta_u < 0.05$. 
In the right panel, we show the prospect of the direct searches at the HL-LHC. As expected, a broader parameter space will be explored if $\zeta_u$ is large.


\section{Multi-Higgs Signatures in the WS Scenario}\label{sec:proposal}

\begin{figure}[t]
\begin{center}
\includegraphics[width=7.2cm]{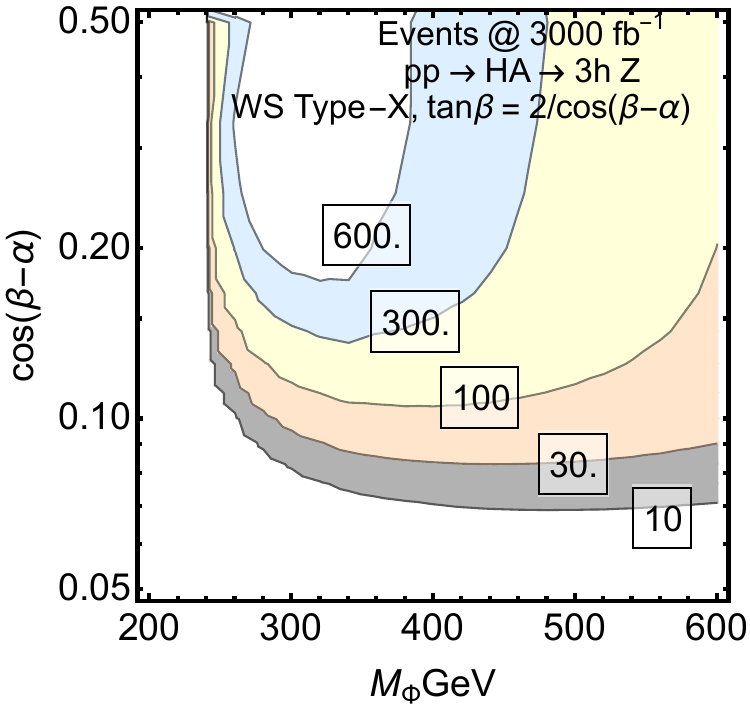}
\includegraphics[width=7.2cm]{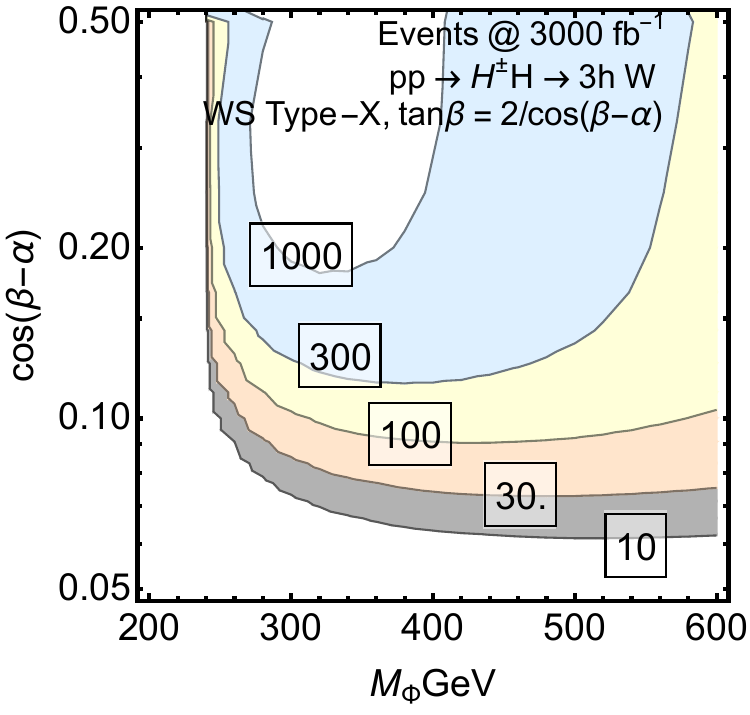}
\caption{New possible search channels and number of events in various multi Higgs final
states at HL-LHC in the Type-X 2HDM}
\label{fig:proposal-X}
\end{center}
\end{figure}

\begin{figure}[t!]
\begin{center}
\includegraphics[width=7.2cm]{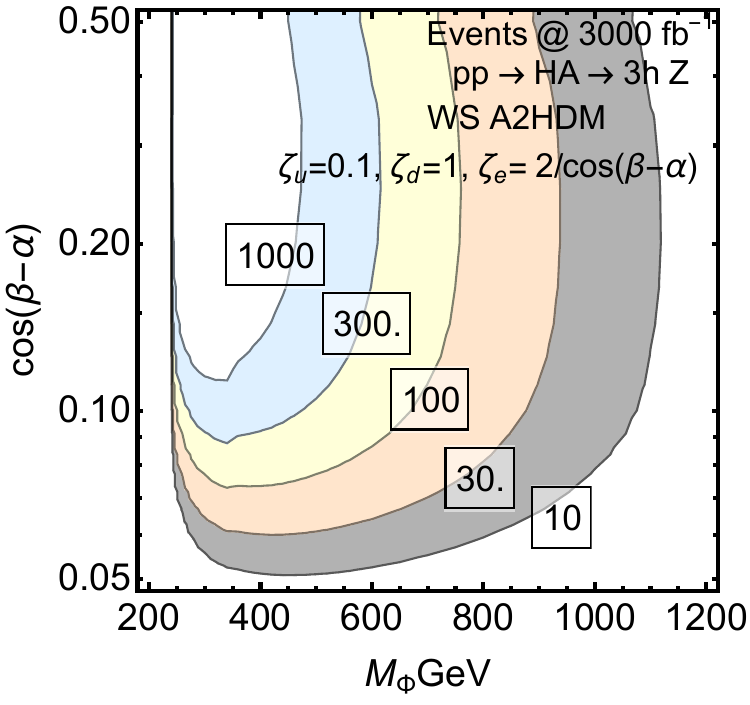}
\includegraphics[width=7.2cm]{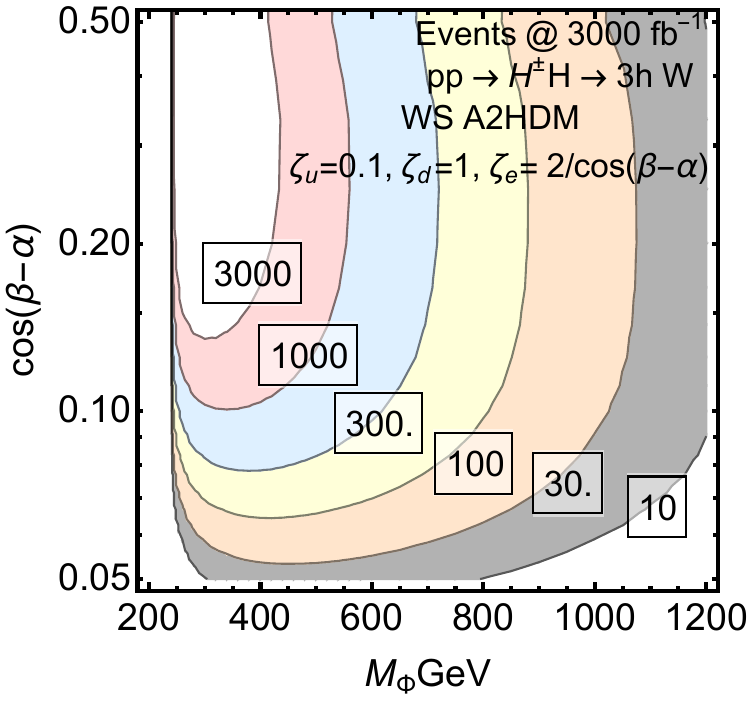}
\caption{New possible search channels and number of events in various multi Higgs final
states at HL-LHC in the A2HDM}
\label{fig:proposal-2HDMA}
\end{center}
\end{figure}

We discuss how we can distinguish the WS scenario from the RS one at the LHC. 
As discussed in Sec.~\ref{sec:comparison}, the additional Higgs bosons predominantly decay into bosonic channels including the discovered Higgs boson $h$ in the WS scenario 
as compared to the RS case. 
Thus, we expect that a larger number of events with multi-Higgs final states can be obtained in the WS scenario than that in the RS case. 

In order to clarify this, we consider a pair production of the additional Higgs bosons via gauge interactions~\cite{Kanemura:2001hz,Cao:2003tr,Belyaev:2006rf,Chun:2018vsn,Bahl:2021str,Mondal:2021bxa}, 
which is quite useful to extract the structure of the Yukawa couplings, because 
its cross section is simply determined by the masses of the additional Higgs bosons in the Higgs alignment limit. 
We here consider the following processes: 
\begin{align}
pp\to Z^*\to AH, \quad pp\to W^{ \pm *}\to H^ \pm  H/ H^ \pm  A, \label{eq:ewprod}
\end{align}
with the succeeding decays
\be
A\to Zh,~~H\to hh/WW/ZZ,~~\textrm{and}~~H^ \pm \to W h. \label{eq:bosonicdecays}
\ee
In the near alignment regime, i.e., $\cos(\beta-\alpha) \simeq 0$, the cross section of the above processes slightly reduces as 
they are proportional to $\sin^2(\beta-\alpha)$ except for the $AH^\pm$ production. 
To estimate the production cross-section we use \texttt{FeynRules}~\cite{Christensen:2008py,Alloul:2013bka} and
$\texttt{MadGraph5\_aMC@NLO}$ \cite{Alwall:2011uj,Alwall:2014hca}. We also use a uniform K-factor of 1.34~\cite{Bahl:2021str}  to incorporate higher-order QCD corrections.
We note that the Yukawa induced productions discussed in the previous section are, of course, useful to 
see the difference between the RS and WS scenarios. 
However, these production cross sections strongly depend on the structure of the Yukawa coupling. 
In particular, when we consider a case with large $\tan\beta$ in the Type-X 2HDM or that with smaller $\zeta_u$ in the A2HDM, 
these processes cannot be used because the cross section is highly suppressed by the smaller top Yukawa coupling for $H$ and $A$. 
%

We estimate the number of events in various multi-Higgs final states with an integrated
luminosity of 3000 fb$^{-1}$.  
For simplicity, we assume a degenerate mass spectrum, i.e., $M_\Phi\equiv M_{H^\pm} (=M_A = M_H)$ as a conservative analysis. 
We note that for the case with a mass difference $M_{H^\pm} > M_H$ and  $M_{H^\pm} = M_A$, the decay branching ratio of $H\to hh$ increases as discussed in Sec.~\ref{sec:comparison}.  
In addition, a portion of the branching ratios of $A\to Zh$ and $H^\pm\to W^\pm h$ is replaced by those of $A\to Z^{(*)}H$ and $H^\pm\to W^{\pm (*)} H$, which eventually can provide a larger 
number of $h$ in the final state due to the succeeding decay of $H \to hh$. 

In Fig.~\ref{fig:proposal-X}, we show the contour plot for the number of events with three Higgs bosons, i.e., 
$hhhZ$ (left) and $hhhW$ (right) as a function of $M_\Phi$ and $\cos(\beta-\alpha)$ in the WS Type-X 2HDM. 
For $\cos(\beta-\alpha)\simeq 0.15$ which is just below the current upper limit by LHC data, 
${\cal O}(100)$ events are expected for the $hhhZ$ mode. 
For the $hhhW$ mode, almost double the number of events is expected as compared with the $hhhZ$ mode. 
We also show the number of events for two Higgs final states in Appendix~\ref{app:2higgs}. 
Similarly in Fig.~\ref{fig:proposal-2HDMA}, we show the number of events for the three Higgs final states in the WS A2HDM. 

\begin{figure}[t]
\begin{center}
\includegraphics[width=7.2cm]{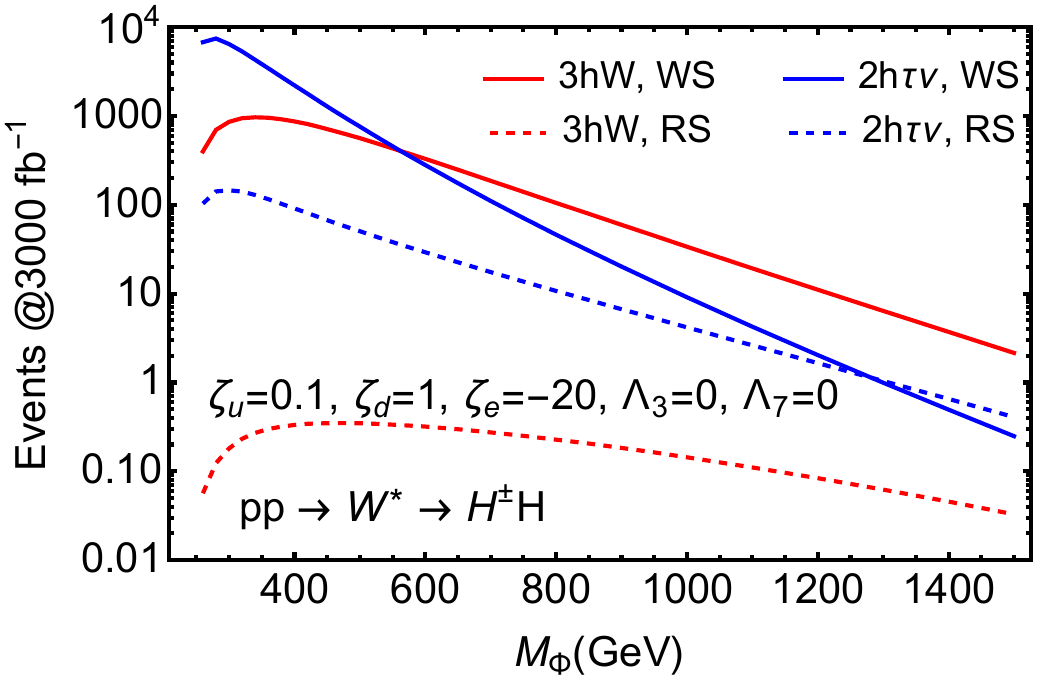}
\includegraphics[width=7.2cm]{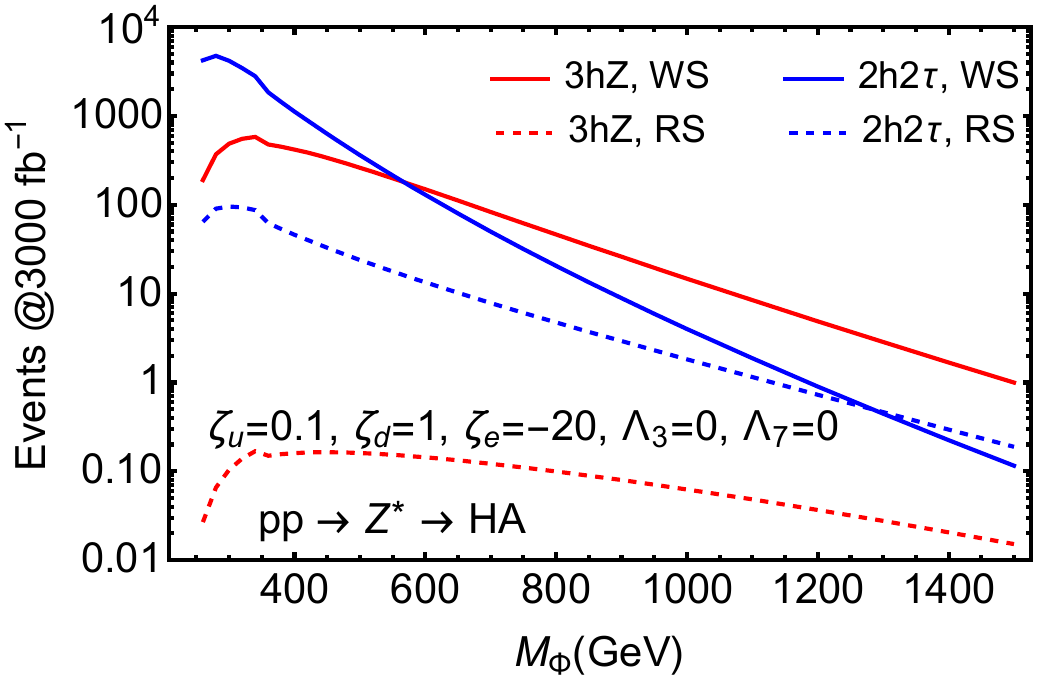}
\caption{Comparison of the number of events in the multi-Higgs final states for the RS (dashed) and WS(solid) cases.}
\label{fig:comparison-events}
\end{center}
\end{figure}

Finally in Fig.~\ref{fig:comparison-events}, we compare the expected number of multi-Higgs events in the A2HDM with the WS and RS scenarios at the HL-LHC. 
It is clear that the event number for the $3h+W(Z)$ final states in the WS case is about three orders of magnitude larger than that in the RS case, and 
almost no event can be seen in the RS case. We have also estimated the $2h+\tau\nu(\tau)$
final state for which the RS event yield is non zero, but remain much smaller than WS case unless the scalars are very heavy. Hence we have considered multi Higgs final state in this section.
Therefore, the measurement of the multi-Higgs final state can be an important probe for testing the WS scenario. 


Before closing this section, let us remark that dedicated signal and background analyses must be required
in order to clarify the significance of these multi-Higgs events, in which we can consider
various final states depending on the decay of the Higgs boson $h$. 
Such a dedicated phenomenological analysis is beyond the scope of the present study.



\section{Conclusion}\label{sec:conclusion}

We have studied the scenario with WS Yukawa couplings for the discovered Higgs boson with the mass of 125 GeV in the Type-X 2HDM and the A2HDM. 
The WS $ht\bar t$ coupling has already been ruled out due to the constraint from measurements of the $h \to \gamma\gamma$ decay at the LHC, while 
WS couplings for down-type quarks and charged leptons are still possible. 
We have focused on the WS scenario for charged leptons in the Type-X 2HDM and the A2HDM under theoretical constraints such as vacuum stability, perturbative unitarity and perturbativity 
as well as the bounds from flavor data, Higgs signal strengths and direct searches at the LHC. 
It has been shown that in the WS scenario, the additional Higgs bosons tend to mainly decay into bosonic states, e.g., $H \to hh/WW/ZZ$, $A \to Zh$ and $H^\pm \to W^\pm h$ as compared with those in the RS scenario. 
We have shown that the existing LHC searches thoroughly explore the
WS Type-X scenario mainly via $H \to hh$, $A\to Zh$ and $A\to \tau\tau$ channels. 
For the A2HDM case, the LHC
limit from the existing searches strongly depends on the up-type alignment parameter
($\zeta_u$). 
For instance, for $\zeta_u = 0.1$, the masses of the additional Higgs bosons up to about 1 TeV can be explored at the HL-LHC. 
Finally, we have proposed to look for a final state of multi-Higgs bosons in association
with gauge bosons via the electroweak production of the additional Higgs bosons. Such multi-Higgs signatures
at the LHC will be an undisputable indication of the WS nature of Yukawa couplings.

\begin{acknowledgments}

This work was supported in part by JSPS KAKENHI Grants No. 20H00160, No. 22F21324 and by Grant-in-Aid for Early-Career Scientists, No.~19K14714. 

\end{acknowledgments}


\appendix

\section{Higgs measurement data}\label{app:higgs-data}

In Sec.~\ref{sec:constraints}, we have performed the $\chi^2$ analysis in order to constrain the parameter space of the 2HDMs, in which 
we have used the current measurements of the Higgs signal strengths for various channels at the LHC. 
For the production, we have taken into account the gluon fusion process (ggF), the vector boson fusion process (VBF), the vector boson associated process ($Vh$) and the top quark associated process ($t\bar{t}h$). 
For the decay, we have taken into account the $h \to \gamma\gamma$, $h \to WW^*$, $h \to ZZ^*$, $h \to \tau\tau$ and $h \to b\bar{b}$ modes. 
In Tab.~\ref{tab:higgs-data}, we summarize the current data that we used in our $\chi^2$ analysis.

 \begin{table}[h!]
 {\footnotesize
\begin{tabular}{|c||c|c|c|c|c|c|}
\hline
& Exp. & $\gamma\gamma$ & $WW^*$ & $ZZ^*$ & $\tau\tau$ & $b\bar b$ \\ \hline
\multirow{2}{*}{ggF}
& A
& $1.03 \pm 0.11$\cite{ATLAS:2020qdt}
& $1.08 \pm 0.19$\cite{ATLAS:2020qdt}
& $0.94 \pm 0.11$\cite{ATLAS:2020qdt}
& $0.95 \pm 0.30$\cite{ATLAS:alltautau}
&\\\cline{2-7}
& C
& $1.07 \pm 0.12$\cite{CMS:allgamgam}
& $1.28 \pm 0.20$\cite{CMS:2020gsy}
& $0.98 \pm 0.12$\cite{CMS:2020gsy}
& $0.97 \pm 0.19$\cite{CMS:tthtautau}
& $2.45 \pm 2.45$\cite{CMS:2020gsy}\\\hline
\multirow{2}{*}{VBF}
& A
& $1.31 \pm 0.25$\cite{ATLAS:2020qdt}
& $0.62 \pm 0.35$\cite{ATLAS:2020qdt}
& $1.25 \pm 0.45$\cite{ATLAS:2020qdt}
& $0.89 \pm 0.18$\cite{ATLAS:alltautau}
& $3.0 \pm 1.6$\cite{ATLAS:2020qdt}\\\cline{2-7}
& C
& $1.04 \pm 0.33$\cite{CMS:allgamgam}
& $0.63 \pm 0.63$\cite{CMS:2020gsy}
& $0.57 \pm 0.41$\cite{CMS:2020gsy}
& $0.68 \pm 0.24$\cite{CMS:tthtautau}
& $1.3 \pm 1.1$\cite{CMS:vbfbb}\\\hline
\multirow{2}{*}{$Vh$}
& A
& $1.32 \pm 0.32$\cite{ATLAS:2020qdt}
& $3.2 \pm 4.3$\cite{ATLAS:2020qdt}
& $1.53 \pm 1.01$\cite{ATLAS:2020qdt}
& $0.95 \pm 0.58$\cite{ATLAS:alltautau}
& $1.02 \pm 0.18$\cite{ATLAS:2020qdt}\\\cline{2-7}
& C
&  $1.34 \pm 0.35$\cite{CMS:allgamgam}
& $1.00 \pm 1.28$\cite{CMS:2020gsy}
& $1.10 \pm 0.85$\cite{CMS:2020gsy}
& $1.80 \pm 0.44$\cite{CMS:tthtautau}
& $1.05 \pm 0.25$\cite{CMS:2020gsy}\\\hline
\multirow{2}{*}{$t\bar t h$}
& A
& $0.9 \pm 0.25$\cite{ATLAS:2020qdt}
&
&
& $1.53 \pm 1.41$\cite{ATLAS:alltautau}
& $0.79 \pm 0.60$\cite{ATLAS:2020qdt}\\\cline{2-7}
& C
& $1.35 \pm 0.34$\cite{CMS:allgamgam}
& $0.93 \pm 0.45$\cite{CMS:2020gsy}
&
& $0.8 \pm 0.7$\cite{CMS:2020gsy}
& $1.13 \pm 0.32$\cite{CMS:2020gsy}\\\hline
\end{tabular}}
\caption{Current data for the measurements of the Higgs signal strengths at the LHC. Here, ``A'' and ``C'' denote ATLAS and CMS, respectively. }
\label{tab:higgs-data}
\end{table}
\section{Number of events for two Higgs final states}\label{app:2higgs}

\begin{figure}[t!]
\begin{center}
\includegraphics[width=4.9cm]{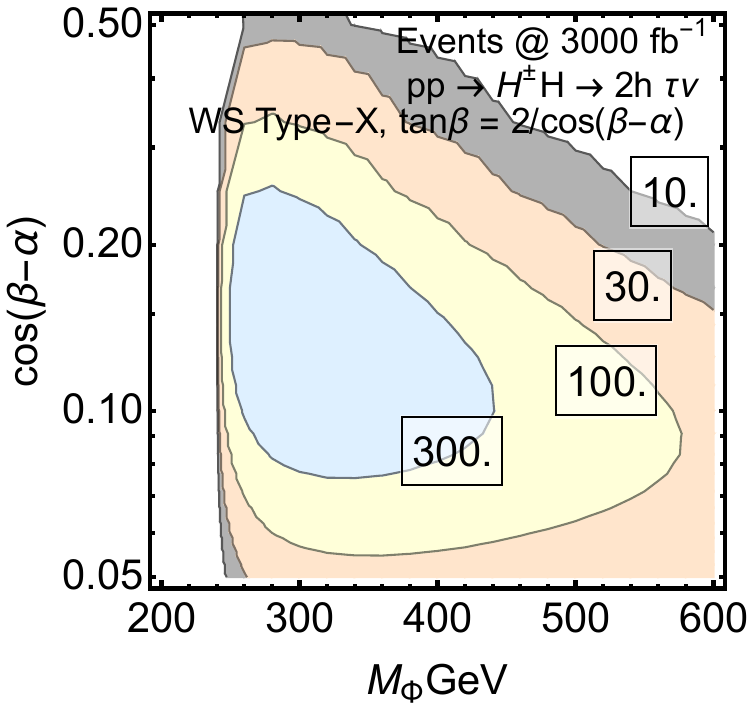}
\includegraphics[width=4.9cm]{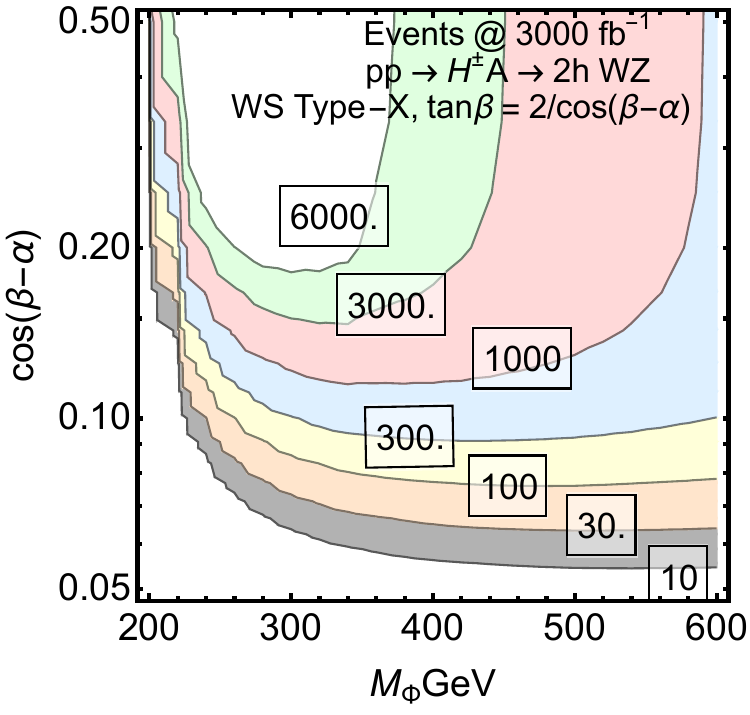}
\includegraphics[width=4.9cm]{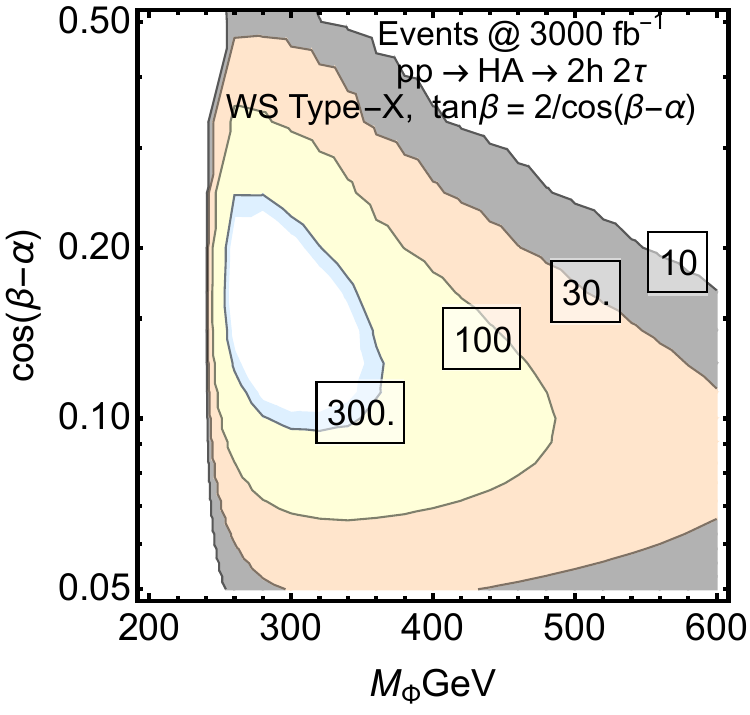}
\caption{New possible search channels and number of events in various multi Higgs final
state at HL-LHC in the WS Type-X 2HDM.}
\label{fig:2higgs_X}
\end{center}
\end{figure}

\begin{figure}[t!]
\begin{center}
\includegraphics[width=4.9cm]{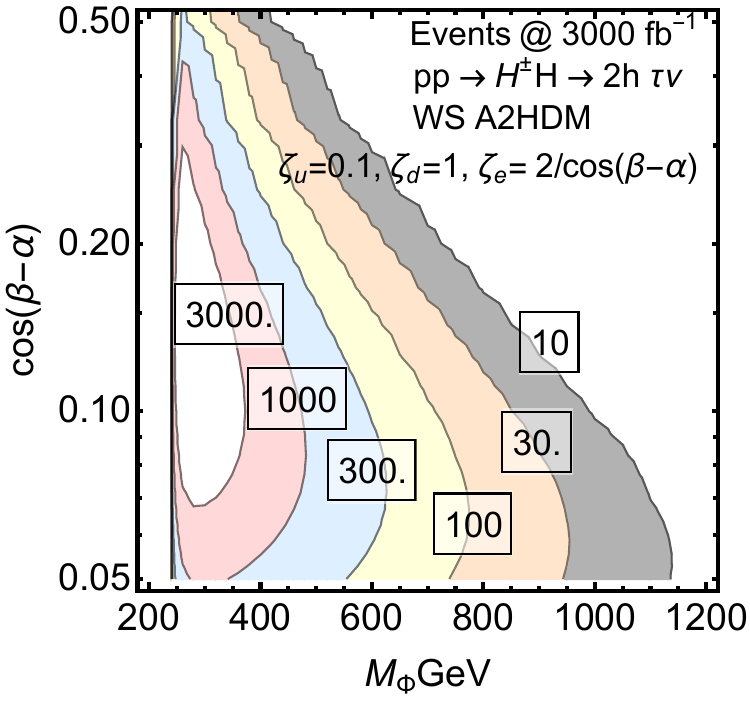}
\includegraphics[width=4.9cm]{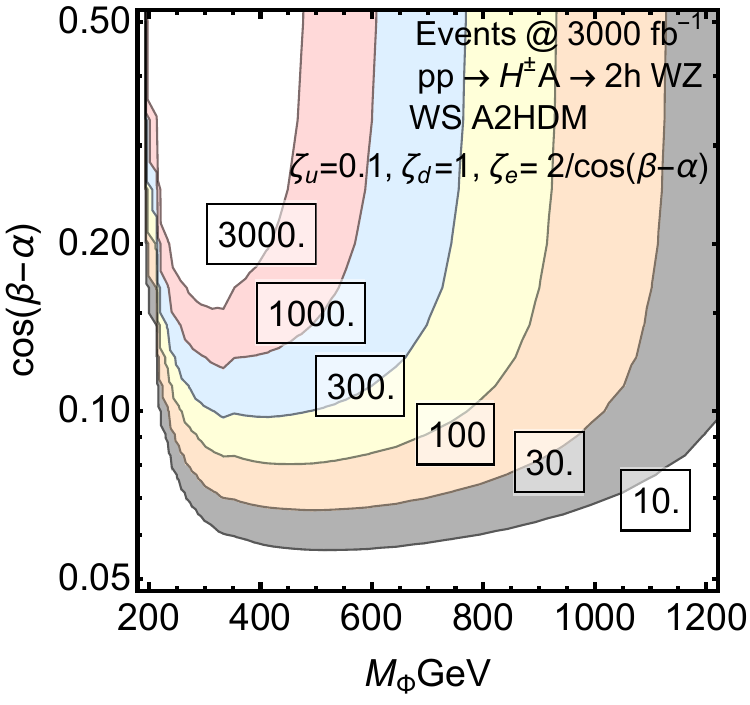}
\includegraphics[width=4.9cm]{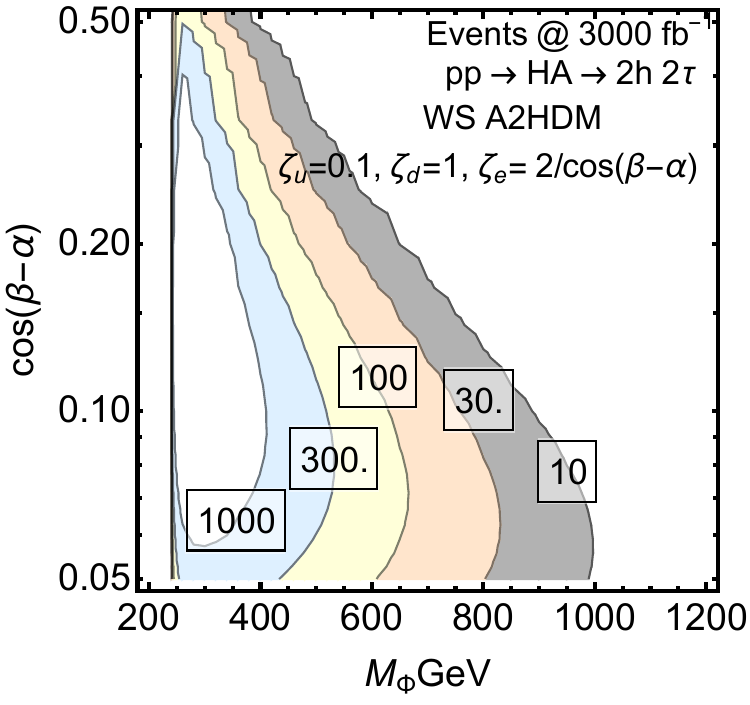}
\caption{New possible search channels and number of events in various multi Higgs final
state at HL-LHC in the WS A2HDM. }
\label{fig:2higgs_A}
\end{center}
\end{figure}

We show the expected number of events for the two Higgs final states from the decay of the additional Higgs bosons. 
In Figs.~\ref{fig:2higgs_X} and \ref{fig:2higgs_A}, we show the contour plots for the number of events in the WS Type-X 2HDM and the WS A2HDM, respectively. 

\bibliographystyle{JHEP}
\bibliography{reference}

\end{document}